\documentclass{article}
\usepackage{arxiv}

\usepackage{cite}
\usepackage{amsmath,amssymb,amsfonts}
\usepackage{textcomp}
\usepackage{algorithmic}
\usepackage{graphicx}
\usepackage{algorithm,algorithmic}
\usepackage{hyperref}
\hypersetup{hidelinks=true}
\hypersetup{colorlinks,allcolors=black,hidelinks=true}
\newcommand{\norm}[1]{\left\lVert#1\right\rVert}

\def\BibTeX{{\rm B\kern-.05em{\sc i\kern-.025em b}\kern-.08em
    T\kern-.1667em\lower.7ex\hbox{E}\kern-.125emX}}
\newtheorem{theorem}{Theorem}
\newtheorem{lemma}[theorem]{Lemma}

\begin{document}
\title{Recursive Quantization for $\mathcal{L}_2$ Stabilization of a Finite Capacity Stochastic Control Loop with Intermittent State Observations}
\author{Shrija Karmakar, and Ritwik Kumar Layek
\thanks{The authors are with Department of Electronics and Electrical Communication Engineering, Indian Institute of Technology, Kharagpur, West Bengal, India (e-mail: ritwik@ece.iitkgp.ac.in).} }

\maketitle

\begin{abstract}
The problem of $\mathcal{L}_2$ stabilization of a state feedback stochastic control loop is investigated under different constraints. The discrete time linear time invariant (LTI) open loop plant is chosen to be unstable. The additive white Gaussian noise is assumed to be stationary. The link between the plant and the controller is assumed to be a finite capacity stationary channel, which puts a constraint on the bit rate of the transmission. Moreover, the state of the plant is observed only intermittently keeping the loop open some of the time. In this manuscript both scalar and vector plants under Bernoulli and Markov intermittence models are investigated. Novel bounds on intermittence parameters are obtained to ensure ${\mathcal{L}_2}$ stability. Moreover, novel recursive quantization algorithms are developed to implement the stabilization scheme under all the constraints. Suitable illustrative examples are provided to elucidate the main results.

\end{abstract}

\section{Introduction}
\label{sec:introduction}
The theory of Linear Quadratic Regulator / Linear Quadratic Gaussian \cite{bellman1954theory,athans1971role,aastrom2012introduction,anderson2007optimal,bertsekas2011dynamic} was developed  to solve the optimal control problem of a linear stochastic system, where a quadratic cost functional of the state and the input is minimized. Here the closed loop system is considered local, and the real valued signals are assumed to flow through the loop. Initially control theory didn't consider the bandwidth related challenges faced by communication engineers. On that front, information theory\cite{shannon1948mathematicalf,cover1999elements,gallager2008principles} was developed to understand the physical limit of point-to-point communication. The accuracy resolutions or the noise processes present in the physical channels impose fundamental limit on the maximum feasible data rate of the channel which is known as `capacity’\cite{shannon1948mathematicalf}. Clearly, an LQG control system with finite capacity channel in loop doesn’t provide simple answers on performance and stability. However, it is a well-known fact now that there exists a duality between stochastic control system and a communication system with feedback \cite{basar2013,charalambous2017information}. In \cite{charalambous2017information}, the authors have demonstrated that stochastic control systems utilize randomized strategies for encoding and transmitting information, similar to Shannon's coding for noisy channels, in order to achieve control-coding capacity. The trade-off between the best achievable data rate through the channel and the best achievable control cost generates considerable interest in recent literature \cite{tanaka2018,nair2007feedback}. It has been shown in \cite{tanaka2018} how Massey's directed information \cite{massey1990causality} between the plant and the controller can be minimized while keeping the quadratic cost within some specified upper bound. In \cite{nair2007feedback}, some interesting results are developed on trade-offs between data rate and different control goals, including stability and asymptotic performances. However, the genesis of the dual optimization of error cost function and information functional lies in rate distortion theory \cite{cover1999elements,shannon1948mathematicalf,tanaka2016semidefinite}. The authors in \cite{tanaka2016semidefinite} investigate Sequential Rate-Distortion (SRD) theory in the context of time-varying Gauss-Markov processes. The study in \cite{you2010minimum} investigates minimum data rates for mean square stabilization of linear systems over lossy digital channels, presenting an innovative method for addressing temporal correlations and fluctuating data rates. For unlimited quadratic cost function, \cite{tanaka2018} shows that the minimum directed information reaches a lower bound which is dependent on the large eigenvalues (outside the unit circle) of the discrete time dynamical system. A similar but more operational result is demonstrated in \cite{kostina2021exact}, where a quantized and encoded state is used to stabilize the control loop. For an unstable scalar linear stochastic system with limited information about system state, the paper establishes the relation between the cardinality of a finite state space and the system gain, which is necessary and sufficient for \(\beta\)-moment stability. The scheme proposed in \cite{kostina2021exact,takashi2022} utilizes a uniform quantizer for data rate efficiency, analyzed using probabilistic and information-theoretic techniques. 

Another practical limitation of a control-communication system lies in sensor or link failure. If the state or output of the plant is not measured due to sensor failure or the transmitted signal over the channel is not received in the estimator or controller, the controller tends to lose its ability to stabilize the loop. A related work in optimal estimation problem is developed in \cite{sinopoli2004kalman}, where the problem of intermittent Kalman filter is introduced. The output is sensed through a random switch which is controlled via a Bernoulli random variable. In \cite{sinopoli2004kalman}, the statistical convergence properties of the estimation error covariance are examined, and a critical intermittence probability is determined at which the estimation error covariance diverges. Upper and lower bounds are provided for the critical intermittence. Several other research works have been pursued in this realm thereafter. The precise critical intermittence probability for bounded error covariance matrix is determined by the known lower bound of the intermittence probability when the observation matrix within the observable subspace is invertible \cite{plarre2009kalman}. The authors of \cite{kluge2010stochastic} have established conditions for bounded estimation error by examining the error behavior of the discrete-time extended Kalman filter in the context of nonlinear systems with intermittent observations. New conditions for the stability of a Kalman filter are introduced in \cite{rohr2014kalman} when intermittent measurements result from communication constraints. The general conditions permit diagonalizable state matrices and finite order Markov processes, thereby expanding the applicability of the current findings to encompass degenerate systems and a wider variety of network models. In \cite{kar2011kalman}, the authors analyses the asymptotic behavior of the random Ricatti equation obtained from the random intermittent switching of the Kalman filter. Similarly,  \cite{kar2012moderate} delves into the dynamics of the random Ricatti equation showing that under certain observability and controllability conditions the distribution functions of the random conditional error covariance matrices converge to an invariant distribution which satisfies a moderate deviation principle. The study in \cite{castro2024regulation} examines the control of a nonlinear system using quantized measurements to accurately follow or reject external signals. It utilizes passivity and output regulation techniques to guarantee the convergence of quantized error and achieve zero asymptotic error in the actual output. \cite{liu2024optimal} examines the optimal control of a linear stochastic system with both additive and multiplicative noise over a finite time horizon, offering a solution for optimal covariance steering, and establishes the existence and uniqueness of the optimal control, along with presenting a result in the form of a matrix Riccati differential equation.

\hspace{1cm}
The efforts to tackle the problem of intermittent observation in a stochastic control loop is ongoing with some success \cite{sinopoli2005lqg}. With packet acknowledgment, the control is linear and follows the separation principle, while critical switching probabilities determine system stability.\cite{li2022lqg} addresses a linear quadratic Gaussian (LQG) tracking problem with safety and reachability constraints when a subset of sensors is subject to an adversarial false data injection attack. A control policy is suggested, which limits the control input by utilizing approximated states that are resolved through the utilization of a quadratically constrained quadratic program. 
In this manuscript, the constraints posed by intermittent observations and finite capacity channel are combined. The contributions of this manuscript are given in section \ref{contributions}.

%\color{black}
\subsection{Contributions}
\label{contributions}
 In this manuscript, the controller is considered to be the infinite horizon linear quadratic Gaussian (LQG) controller. When the observation switch is `ON', the state is measured, quantized, and transmitted to the controller to generate the control signal, which will make the closed loop stable. However, when the observation switch is `OFF', the state is not measured and the loop is open and unstable. The main contributions of this manuscript are twofold. Firstly, the necessary and sufficient conditions on the intermittence parameters for $\mathcal{L}_2$ stability of the overall system are given. Secondly, when the intermittence conditions are met, a recursive quantization algorithm is designed to transmit the state through the finite capacity channel. Both of these contributions are novel and they extend several works from the contemporary literature as discussed in section \ref{sec:introduction}. 
  
\subsection{Organization of the paper}
This paper is organized as follows. Section \ref{sec:introduction} introduces the problem with a survey of the relevant literature along with the main contribution of the manuscript, and some useful notations. In section \ref{formulation}, the mathematical description of the problem is introduced. The main theoretical contributions of the manuscript are given in section \ref{results}. Illustrating examples, and simulations are shown in section \ref{simulation}. Few concluding remarks are given in section \ref{conclusion}. The proofs of all the lemmas, and theorems are presented in section \ref{appendix}.

\subsection{Notations}
Some useful notations, terminologies, and concepts are mentioned here.
\begin{itemize}
    \item  The real line is denoted by $\mathbb{R}$. All logarithms in this manuscript are assumed to have base $2$. $\delta(k)$ is the Kronecker's delta function. Matrices are written in capital letters. Vectors and scalars are written using small letters. For a vector $x$, the $\mathcal{L}_2$ norm is denoted by $\norm{x}_2$. $A'$ is the transpose of $A$. The determinant of matrix $A$ is denoted by $|A|$. A zero vector of size $(n\times 1)$ is denoted by $0_n$. A zero matrix of size $(n\times m)$ is denoted by ${\bf 0_{n \times m}}$. For a $(n\times n)$ matrix $A$, $(n\times 1)$ vector $v$, and scalar (real or complex) $\lambda$, if the following equation is satisfied: $Av = \lambda v$, then $\lambda$, and $v$ are called eigenvalue, and eigenvector of the matrix $A$ respectively.   
    $A \succ B$ means $\left (A-B\right )$ is a positive definite matrix. $A \succeq B$ means $\left (A-B\right )$ is a positive semi-definite matrix. The set of all $n\times n$ positive definite matrices is denoted by $M_n^+$. The set of all $n\times n$ positive semi-definite matrices is denoted by $M_n$.
    For two matrices $A$, and $B$, Minkowskii's inequality of determinants is given by: $|A+B|^{\frac{1}{n}} \ge |A|^{\frac{1}{n}} + |B|^{\frac{1}{n}}$. A set $C$ is called a cone if for every point $x \in C$ and for all $\theta \ge 0$, $\theta x \in C$. A cone is called a convex cone if for any $x_1,x_2 \in C$ and $\theta \in [0,1]$, $\theta x_1 + (1-\theta) x_2 \in C$. For a linear map $\mathcal{T} : \mathbb{R}^n\rightarrow \mathbb{R}^m$, the spectral norm \cite{horn2012matrix} of $\mathcal{T}$ is defined as: $\norm{\mathcal{T}}_2 = \inf \{c\ge0 : \norm{\mathcal{T}(x)}_2 \le c\norm{x}_2 \forall x \in \mathbb{R}^n \}$. An operator $\mathcal{T} : \mathbb{R}^n \rightarrow \mathbb{R}^m$ is called a contraction operator if $\norm{\mathcal{T}}_2 < 1$.
           
    \item A triple $\left(\Omega,\mathcal{B},\mathbb{P}\right)$ is called a probability space where $\Omega$ is the sample space, $\mathcal{B}$ is the $\sigma$-algebra, and $\mathbb{P}$ is the probability measure. The map ${\bf x}: \left(\Omega,\mathcal{B},\mathbb{P}\right)\rightarrow \left(\mathbb{R},\mathcal{B}(\mathcal{R}), F_{\bf x}\right)$ is called a random variable, where the domain space is some underlying probability space, and the range space is the induced measure space of the real line. $\mathcal{B}(\mathcal{R})$ is the Borel $\sigma$-algebra of the real line, and $F_{\bf x}$ is the probability distribution function of $\bf x$ characterizing the induced measure. In this manuscript probability is denoted by $P$. Here, $F_{\bf x}(\xi) = P({\bf x}\le\xi) = \mathbb{P}\left({\bf x}^{-1}\left(\left(-\infty, \xi\right]\right)\right)$. Random variables/vectors are written using bold small letters. For notational simplicity we will not mention the probability spaces explicitly everywhere. The standard measure of the real line is the Lebesgue measure. $\mathcal{U}$, $\mathcal{B}ern$, and $\mathcal{N}$ denote the uniform distribution, Bernoulli distribution, and Gaussian distribution, respectively. $\mathbb{E}$ is the expectation operator defined by the Stieltjes integral: $\mathbb{E}({\bf x}) = \int_{\mathbb{R}}\xi dF_{\bf x}(\xi)$. The variance of a random variable ${\bf x}$ is defined by: $\sigma_{\bf x}^2 = Var({\bf x}) = \mathbb{E}({\bf x}-\mathbb{E}({\bf x}))^2$. For a indexed random variable ${\bf x}_k$, the variance is denoted by the shorthand notation $\sigma_k^2$. If $\bf x$ is a Gaussian random variable with mean $\mu$, and variance $\sigma^2$, it is denoted as: ${\bf x}\sim\mathcal{N}(\mu,\sigma^2)$. A random vector ${\bf x}$ of size $(n\times 1)$ is an $n$-tuple of random variables. The covariance matrix of a random vector ${\bf x}$ is defined as: $P_{\bf x}=Cov({\bf x}) = \mathbb{E}\left(\left({\bf x}-\mathbb{E}({\bf x})\right)\left({\bf x}-\mathbb{E}({\bf x})\right)'\right)$. Again, for indexed random vector ${\bf x}_k$, the covariance matrix is denoted by the shorthand notation $P_k$. ${\bf x}\sim\mathcal{N}(\mu,P_{\bf x})$ denotes a Gaussian random vector with mean vector $\mu$ and covariance matrix $P_{\bf x}$.  ${\bf x}_k$ is a random variable/vector at time $k$. $x_k$ is the scalar/vector value at time $k$. Unless stated otherwise, all random variables/vectors used in this manuscript are assumed to be of zero mean.
    A sequence of random variables/vectors is denoted by: ${\bf x}_{0:n} = \{{\bf x}_0,{\bf x}_1, \dots {\bf x}_n\}$. The conditional expectation is defined as:
   $\mathbb{E}\left({\bf x}|{\bf y}\right) = \int_{\mathbb{R}_{\bf x}} x dF_{{\bf x}|{\bf y}}(x|{\bf y})$. Here, $\mathbb{R}_{\bf x}$ denotes the range space of ${\bf x}$. Clearly $\mathbb{E}\left({\bf x}|{\bf y}\right)$ is a random variable. The law of total expectation suggests that $\mathbb{E}\left({\bf x}\right)= \mathbb{E}\left(\mathbb{E}\left({\bf x}|{\bf y}\right)\right)$. Conditional variance of the random variable ${\bf x}_k$ conditioned on the sequence of random variables ${\pmb \gamma}_{0:k-1}$ is denoted by: ${\pmb\sigma}_{k|{\pmb\gamma}_{0:k-1}}^2 = \mathbb{E}({\bf x}_k^2|{\pmb\gamma}_{0:k-1})$. Conditional covariance matrix of the vector ${\bf x}_k$ conditioned on the sequence of random variables ${\pmb\gamma}_{0:k-1}$ is denoted by: ${\bf P}_{k|{\pmb\gamma}_{0:k-1}}=\mathbb{E}({\bf x}_k{\bf x}'_k|{\pmb\gamma}_{0:k-1})$. For zero mean random variables, a Hilbert space of random variables can be constructed by defining the inner product as: $<{\bf x}, {\bf y}> = Cov({\bf x}, {\bf y})$. Hence, the $\mathcal{L}_2$ norm is given by: $\norm{\bf x}_2 = \sqrt{Var({\bf x})} = \sigma_{\bf x}$. A sequence of random variables ${\bf x}_{0:\infty}$ is said to be $\mathcal{L}_2$ convergent to ${\bf y}$ if $\lim_{n\rightarrow\infty} \norm{{\bf x}_n-{\bf y}}_2 = 0$. 
    
 \item The differential entropy of a continuous random variable ${\bf x}$ with a probability density function $f_{\bf x}\left(x\right)$ is defined as $h\left({\bf x}\right) = \mathbb{E}\left[-log\left(f_{\bf x}\left({\bf x}\right)\right)\right] = -\int_{\mathbb{R}_{\bf x}}log\left(f_{\bf x}\left(x\right)\right) f_{\bf x}\left(x\right) dx$. The entropy of a discrete random variable ${\bf z}$ with probability mass function $p_{\bf z}(z)$ is given by $H({\bf z})=-\Sigma_{z\in supp\{{\bf z}\}}p_{\bf z}(z) log\left(p_{\bf z}(z)\right)$. Quantization is the process of countably partitioning the sample space of a continuous random variable and assigning a value for every partition to generate a discrete random variable. If a continuous random variable ${\bf x}$ with differential entropy $h({\bf x})$ is quantized uniformly using the quantization step-size $\Delta$ (Lebesgue measure of the partition over the range of {\bf x}), the entropy $H({\bf x}^{\Delta})$ of the quantized discrete random variable ${\bf x}^{\Delta}$ is given by $H({\bf x}^{\Delta}) = h({\bf x}) - log(\Delta)$. The typical support set of a random variable ${\bf x}$ is a set $S\subseteq \mathbb{R}$ such that $P({\bf x}\in S) \ge 1-\epsilon$ for some arbitrarily small $\epsilon > 0$.
\item A discrete time LTI autonomous system $x_{k+1}=Ax_{k}$, $x_k \in \mathbb{R}^n$, $k = 0,1,2,\dots$ is called asymptotically stable if and only if there exists a unique solution $P \succ 0_{n\times n}$ for the Lyapunov equation $APA' - P + Q = 0_{n\times n}$, where $Q \succ 0_{n\times n}$.

\end{itemize}

\section{Problem Formulation} 
\label{formulation}
Consider the discrete time LTI system
\begin{eqnarray}
\label{vector_plant}
{\bf x}_{k+1}= A{\bf x}_{k}+ B{\bf u}_{k}+{\bf w}_{k},\hspace{4mm} k=0,1,2,\dots
\end{eqnarray}
where ${\bf x}_k\in \mathbb{R}^n$ is the state vector, ${\bf u}_k\in\mathbb{R}^m$ is the input vector with $m<n$, and ${\bf w}_k\in\mathbb{R}^n$ is the additive white gaussian noise vector. The matrix ${ A}$ is assumed to be diagonalizable, and the system is $\left({ A},{ B}\right)$ controllable. The initial state ${\bf x}_0\sim \mathcal{N}\left(0,{ P}_{0}\right)$, ${P}_{0}\succ {\bf 0}_{n\times n}$, the noise process ${\bf w}_{k}\sim \mathcal{N}\left(0,W\right)$, $W\succ {\bf 0}_{n \times n}$, and $\mathbb{E}\left({\bf w}_{k}{\bf w}_{j}^{\top}\right)=W\delta(k-j)$. It is also assumed that $\mathbb{E}\left({\bf x}_{k}{\bf w}_{j}^{\top}\right)={\bf 0_{n\times n}}$ $\forall j\ge k$.

\begin{figure}[h]
\centering
\includegraphics[width=3.5in]{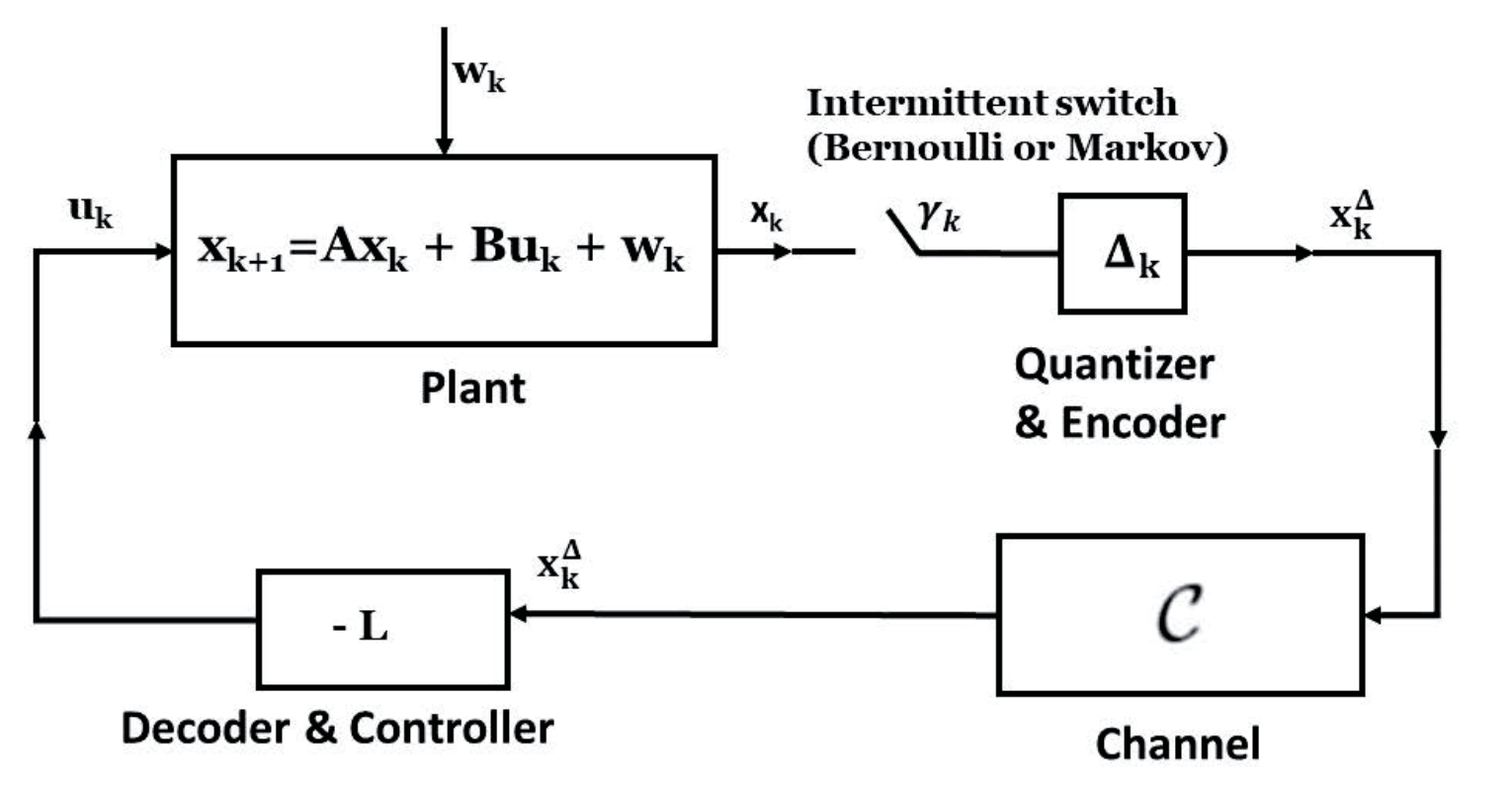}
\caption{The finite capacity LQG loop with intermittent observations}
\label{loopLQG}
\end{figure}

In infinite horizon LQG control synthesis, the objective is to design the optimal control policy which minimizes the expected quadratic cost function:
\begin{eqnarray}
J({\bf x}_{0:\infty}, {\bf u}_{0:\infty}) = \sum_{k =0}^{\infty }\mathbb{E}\left({\bf x}_{k}^{\top}Q{\bf x}_{k}+{\bf u}_{k}^{\top}R{\bf u}_{k}\right)
\end{eqnarray}
Here the costs of the state and the input are given by the matrices $Q \succ {\bf 0}_{n\times n}$ and $R \succ {\bf 0}_{m \times m}$. In state feedback optimal control it is assumed that the state ${\bf x}_k$ is accessible to the controller to synthesize optimal control law which takes the linear form of ${\bf u}_k = -L{\bf x}_k$, where $L$ is the optimal gain matrix. However, in the control loop shown in Fig. \ref{loopLQG}, there exists a channel with finite capacity ($\mathcal{C}$) between the plant and the controller. 
Hence, the observation of real valued state is no longer feasible. For error free communication through the channel, the state signal at the plant needs to be quantized and encoded to keep the data rate below the capacity value $\mathcal{C}$ \cite{shannon1948mathematicalf}. \\
 Furthermore, due to possible link failure and packet drops, the communication link is modeled with a random switch which controls the transmission through the channel leading to intermittent observations of the quantized state at the controller. Two different models of  intermittence is considered, namely the Bernoulli switch, and the Markov switch. 

%
% We consider the observation of the state $x_k$ at time $k$ as a Bernoulli random variable $\gamma$ with $P(\gamma =1) = p$ as the probability of the switch `OFF' state. The block diagram of the system is shown in Fig. \ref{LQG}. 
The block diagram of the entire system is shown in Fig. \ref{loopLQG}. The plant is the dynamical system which generates the state according to Equation. \ref{vector_plant}. The intermittent switch is controlled by a binary random variable whose `0' state implies the switch is `ON' and the state signal can be transmitted to the controller closing the loop. On the other hand, switch state `1' implies the switch is `OFF' and the signal cannot be further processed which will eventually make the loop open. The link failure event is rather rarer and the corresponding switching probabilities will be studied. This will make further discussion more convenient. The mathematical description and relevant derivations are given in section \ref{results}. Sensing the intermittence, the quantizer-encoder block will update the quantization step-size such that the capacity constraint is satisfied. If the switch is `ON', the state will be quantized and encoded before sending it through the channel. The decoder-controller block will receive the signal and construct the linear state feedback optimal control law using two pieces namely, the gain matrix computed from the infinite horizon linear quadratic regulator analysis, and the quantized state information. This control law will close the loop driving the system towards stability.The analysis with a scalar plant is introduced first. The objective will be to find corresponding bounds of the intermittence parameters such that the stochastic system is $\mathcal{L}_2$ stabilized. 

%Furthermore, we assume that the observation of the state is intermittent, which is controlled by a random switch. The block diagram of the system is shown in Fig. \ref{iLQG}. Henceforth, two different scenarios are addressed. 
%\begin{enumerate}
%\item We consider the observation of the state $x_k$ at time $k$ as a Bernoulli random variable $\gamma$ with $P(\gamma =1) = p$ as the probability of the switch `OFF' state. 
%\item We consider the observation of the state $x_k$ at time $k$ as a Markov process with the random variable $\gamma = 1$ indicating the switch 'OFF' state. The state transition matrix of the Markov process is given by:\\\\
%$T_0$ =\[ \begin{bmatrix}
%    1-p & p \\
%    q & 1-q \\
%    \end{bmatrix}\] \\\\
%$p$ is the probability of transition from $\gamma = 0$ to $\gamma = 1$, and $q$ is the probability of transition from $\gamma = 1$ to $\gamma = 0$. For a generalized scenario, it can be considered that $p \ne q$.
%The block diagram of the system is shown in Fig. \ref{LQG2}. 
%\end{enumerate}

\section{Main Results}
\label{results}
\subsection{Bernoulli intermittence model}
\subsubsection{Scalar Plant}
\label{scalar_bernoulli}
The scalar version of the system introduced in section \ref{formulation} is given below. 
\begin{eqnarray}
{\bf x}_{k+1}= \alpha {\bf x}_{k}+ b {\bf u}_{k}+{\bf w}_{k},\hspace{4mm} k=0,1,2,\dots
\label{scalar_plant}
\end{eqnarray}
The initial state is represented as ${\bf x}_0 \sim \mathcal{N}\left(0,\sigma_0 ^2\right)$ where $\sigma_0^{2}$ is the variance of the initial state. 
An important assumption of the optimal control design problem is $|\alpha| > 1$ and also, the control applied on the system is infinite horizon LQR. The scalar noise process is $w_k \sim \mathcal{N}\left(0,\sigma_{\bf w} ^2\right)$ where $\sigma_{\bf w}^{2}$ is the variance of the noise. Moreover, $\mathbb{E}\left({\bf w}_k {\bf w}_l\right)=0  \forall k\ne l$, and $\mathbb{E}\left({\bf x}_k {\bf w}_l\right)=0  \forall l\ge k$. The observed state sequence is up to time $k$ denoted by: $x_{0:k} = \{x_0,x_1,\dots,x_k\}$. The realized input sequence up to time $k$ is denoted by: $u_{0:k} = \{u_0,u_1,\dots,u_k\}$. The random sequences are denoted likewise. The intermittence and capacity constraints are briefly introduced in section \ref{formulation}.
The state of the intermittent switch at time $k$ is the the random variable ${\pmb \gamma}_k \sim  \mathcal{B}ern(p)$, $0\le p\le 1$. $P({\pmb\gamma}_k =1)=p$ is the probability of the switch being open at time $k$, and then the state ${\bf x}_k$ cannot be observed and transmitted to the decoder-controller block. The Bernoulli intermittence sequence up to time $k$ is given by: ${\pmb\gamma}_{0:k} = \{{\pmb \gamma}_0, {\pmb \gamma}_1,\dots,{\pmb \gamma}_k\}$. The capacity of the channel between the quantizer-encoder and the decoder-controller is $\mathcal{C}$. 
The necessary and sufficient conditions for $\mathcal{L}_2$ stabilizability of the system are discussed in the following lemmas and theorems. 
\begin{lemma}
\label{lemma1}
    For the scalar plant described in section \ref{scalar_bernoulli}, in absence of quantization (i.e., assuming $\mathcal{C}\rightarrow\infty$), the conditional state variance at the $(k+1)^{th}$ time step is given by:
\begin{eqnarray}
    {\pmb \sigma}_{k+1|{\pmb \gamma}_{0:k}}^2 = \left(\prod_{j=0}^{k} {\pmb \xi}_{j}^2 \right)\sigma_{0}^2 + \left[\sum_{j=0}^{k-1}\left( \prod_{i=j+1}^{k} {\pmb \xi}_{i}^2 \right) + 1 \right] \sigma_{\bf w}^2
\end{eqnarray},
    and the (expected) state variance at the $(k+1)^{th}$ time step is given by:
    \begin{eqnarray}
    \label{lemma1_eqn}
        \sigma_{k+1}^2 = \omega^{2(k+1)}\sigma_{0}^2 + \frac{1 - \omega^{2(k+1)}}{1 - \omega^2}\sigma_{\bf w}^2
    \end{eqnarray}
    Here, ${\pmb \xi}_k = \alpha {\pmb \gamma}_k + (\alpha - bl)(1 - {\pmb \gamma}_k)$ , $l$ is the infinite horizon LQR gain, and $\omega^2 = \alpha^2 p+(\alpha-bl)^2(1-p)$. 
    
\end{lemma}

{\bf Proof}:    The proof is provided in section \ref{proof1}.

The scenario will be a little different when the finite capacity channel is considered between the switch and the controller. The random state ${\bf x}_k$, if observed will be quantized and encoded such that the signal data rate is below the capacity of the channel to ensure asymptotic error free detection of the quantized signal at the detector-controller block. For any random map whose range is not of finite support, finite quantization can be done using the notion of `typicality' of the range space. For a zero mean Gaussian random variable ${\bf x}_k$, the `typical range' set can be assumed to be $\mathcal{S}_{{\bf x}_k}=\left[-2^{\left(h\left({\bf x}_k\right)-1\right)}, +2^{\left(h\left({\bf x}_k\right)-1\right)}\right]$. Here, ${\bf x}_k \sim \mathcal{N}\left(0,\sigma_k ^2\right)$, and $h\left({\bf x}_k\right) = \frac{1}{2}\log \left(2\pi e \sigma_k^2\right)$ is the differential entropy of ${\bf x}_k$. Another popular choice of typicality set is of course the $6\sigma$ support set, which is in fact a little bigger than the set derived from differential entropy. However, to derive the bounds analytically, the typical set is used in this manuscript. Suppose, the quantization step size at time $k$ is ${\pmb\Delta}_k$, which is obviously a random variable. Denote, $\Delta_k=\mathbb{E}({\pmb\Delta}_k)$ In this manuscript, uniform quantization is considered. In this setup, the one step state variance update equation is stated in Lemma \ref{lemma2}. 

%************************************

\begin{lemma}
\label{lemma2}
    For the system discussed in section \ref{scalar_bernoulli}, if $\mathcal{C} < \infty$, and if the quantization step-size at the $k^{th}$ time step is ${\bf \Delta}_k$, conditioning on the intermittence random sequence, the one step conditional state variance update equation at the $(k+1)^{th}$ time step is given by:
    \begin{eqnarray}
    \label{cond_var_bern_scalar_eqn}
        {\pmb\sigma}_{k+1|{\pmb\gamma}_{0:k}}^2={\pmb\xi}_k^2{\pmb\sigma}_{k|{\pmb\gamma}_{0:k-1}}^2+b^2 l^2(1-{\pmb\gamma}_k)\frac{{\pmb\Delta}_{k|{\pmb\gamma}_{0:k-1}}^2}{12}+\sigma_{\bf w}^2
        \end{eqnarray}
, and the one step (expected) state variance update equation at the $(k+1)^{th}$ time step is given by: 
    \begin{eqnarray}
    \label{var_bern_scal_eqn}
        \sigma_{k+1}^2 = \omega^2\sigma_{k}^2 + b^2l^2(1-p)\frac{\Delta_{k}^2}{12} + \sigma_{\bf w}^2
    \end{eqnarray}
\end{lemma}

{\bf Proof}:  The proof is provided in section \ref{proof2}.

% \begin{figure}[h]
% \centering
% \includegraphics[width=3.5in]{gaussian.eps}
% \caption{The conditional density of the of the state random variable conditioned on a quantization region}
% \label{iLQG}
% \end{figure}

%****************************

\begin{lemma}
\label{lemma3}
    For the system discussed in section \ref{scalar_bernoulli}, the one step conditional quantization step size update equation for the $(k+1)^{th}$ time step is given by:
    \begin{eqnarray}
    \label{scal_bern_cond_quant_eq}
     {\pmb \Delta}_{k+1|{\pmb \gamma}_{0:k}}^2 ={\pmb \Delta}_{k|{\pmb \gamma}_{0:k-1}}^2\left[{\pmb \xi}_k^2 + \eta G(1-{\pmb\gamma}_k)\right] + \eta\sigma_{\bf w}^2 + \epsilon(1 - {\pmb \xi}_k^2)
    \end{eqnarray}
, and the one step (expected) quantization step size update equation for the $(k+1)^{th}$ time step is given by:
    \begin{eqnarray}
    \label{scal_bern_quant_eq}
        \Delta_{k+1}^2 = \Delta_{k}^2\left[\omega^2 + \eta G(1-p)\right] + \eta\sigma_{\bf w}^2 + \epsilon(1 - \omega^2)
    \end{eqnarray}.
    
    Here, $\eta = \frac{2\pi e} {2^{2\mathcal{C}}}$, $G = \frac{b^2 l^2}{12}$, and $\epsilon>0$ is a small constant. 
\end{lemma}

{\bf Proof}:  The proof is provided in section \ref{proof3}.

Based on these lemmas, the result on intermittence probability upper bound for the Bernoulli model is stated in the next theorem. 
\begin{theorem}
\label{theorem4}
For the system discussed in section \ref{scalar_bernoulli}, the necessary and sufficient conditions for $\mathcal{L}_2$ stabilization are given below. 
    \begin{enumerate}
        \item  For $\mathcal{C} \rightarrow \infty$, the necessary and sufficient condition for $\mathcal{L}_2$ stabilizability of the system is given by:
        \begin{eqnarray}
        \label{scal_bern_iff_inf}
         p < \frac{1 - (\alpha-bl)^2 }{\alpha^2 - (\alpha-bl)^2}    
        \end{eqnarray}
    , and if the condition is satisfied, the state variance asymptotically converges as follows:
        \begin{eqnarray}
        \label{scal_bern_iff_inf_var}
           \lim_{n\rightarrow
           \infty} \sigma_n^2 = \left(\frac{\sigma_{\bf w}^2 }{1 - \omega^2 }\right)
        \end{eqnarray}.
        \item For $\frac{1}{2}\log\left(\frac{\pi e \alpha^2}{6}\right)<\mathcal{C} < \infty$, the necessary and sufficient condition for $\mathcal{L}_2$ stabilizability of the system is given by:
        \begin{eqnarray}
        \label{scal_bern_iff_fin}
         p < \frac{1 - (\alpha-bl)^2 - \eta G}{\alpha^2 - (\alpha-bl)^2 -  \eta G}    
        \end{eqnarray}
        %\left(\frac{\pi e b^2 l^2}{6}\right)2^{2\mathcal{C}}
        , and if the condition is satisfied, the state variance asymptotically converges as follows:
        \begin{eqnarray}
        \label{scal_bern_iff_fin_var}
           \lim_{n\rightarrow
           \infty} \sigma_n^2 = \left(\frac{\sigma_{\bf w}^2 + G(1-p)\epsilon}{1 - \omega^2 - \eta G(1-p)}\right)
        \end{eqnarray}.
        Here, $\eta = \frac{2\pi e} {2^{2\mathcal{C}}}$, $G = \frac{b^2 l^2}{12}$, and $\epsilon>0$ is a small constant. 
    
        \item For $\mathcal{C}\le \frac{1}{2}\log\left(\frac{\pi e \alpha^2}{6}\right)$, the system is always $\mathcal{L}_2$ unstable. 
    \end{enumerate}

\end{theorem}

{\bf Proof}:  The proof is provided in section \ref{proof4}.

\subsubsection{Vector Plant}
\label{vect_bern_sys}
The details of the vector plant is already sketched in section \ref{formulation}. The Bernoulli switching is already introduced in section \ref{scalar_bernoulli}. The following lemmas will help to develop the main theorem for the vector plant under Bernoulli intermittence.  
\begin{lemma}
\label{lemma5}
\label{vector_bernoulli_covariance}
For the vector plant described in section \ref{vect_bern_sys}, the conditional state covariance matrix in absence of quantization, (i.e, when the channel capacity $(\mathcal{C})$ is infinite) at the $(k+1)^{th}$ time step is given by:
\begin{eqnarray}
\label{vec_bern_cond_cov_eq}
   \nonumber {\bf P}_{k+1|{\pmb \gamma}_{0:k}}&=&\mathbb{E}\left({\bf x}_{k+1}{\bf x}'_{k+1}|{\pmb \gamma}_{0:k}\right)\\
     &=&\left(\prod_{j=k}^{0}{\pmb\Lambda}_j\right) {P}_0\left(\prod_{j=0}^{k}{\pmb\Lambda}'_j\right)
    +\left[\sum_{j=0}^{k-1}\left\{\left(\prod_{i=k}^{j+1}{\pmb\Lambda}_i\right)W\left(\prod_{i=j+1}^{k}{\pmb\Lambda}'_i\right)\right\}+W\right]
\end{eqnarray}
Where, ${\bf P}_0 = \mathbb{E}\left({\bf x}_{0}{\bf x}'_{0}\right)$, and ${\pmb\Lambda}_k = {\pmb\gamma}_k A + \left(1-{\pmb\gamma}_k\right)\left(A-BL\right)$.
\end{lemma}
{\bf Proof}:  The proof is provided in section \ref{proof5}.

%&&&&&&&&&&&&&&&&&&&&&&&&&&&&  RKL
\begin{lemma}
\label{lemma6}
\label{vector_bernoulli_covariance_expected}
For the vector plant described in section \ref{vect_bern_sys}, the one step dynamics of the state covariance matrix ($P_k$) in absence of quantization, (i.e, when the channel capacity $(\mathcal{C})$ is infinite) is given by:
\begin{eqnarray}
\label{vec_bern_cov_eq}
    P_{k+1} = pAP_{k}A'+(1-p)(A-BL)P_{k}(A-BL)'+W
\end{eqnarray}
where, $P_k = \mathbb{E}\left({\bf x}_k {\bf x}'_k\right)$. \\
\end{lemma}
{\bf Proof}:  The proof is provided in section \ref{proof6}.

Next, the main theorem for $\mathcal{L}_2$ stabilization of the vector plant without quantization under Bernoulli intermittence can be stated. 
\begin{theorem}
\label{theorem7}
\label{vector_bernoulli_probability}
   For the vector plant described in section \ref{vect_bern_sys}, if capacity is infinite $(\mathcal{C}\rightarrow \infty)$, the necessary and sufficient condition on the intermittence probability ($p$) for the asymptotic state covariance matrix to be bounded, and unique is given by:
\begin{eqnarray}
 p< \frac{1-\norm{A-BL}_2^2}{\norm{A}_2^2-\norm{A-BL}_2^2}
 \end{eqnarray}
 Where, $\norm{A}_2$ is the spectral/operator norm of the matrix $A$.
\end{theorem}

{\bf Proof}:  The proof is provided in section \ref{proof7}.

However, one important point to mention is that the condition of Theorem \ref{theorem7} will be satisfied if and only if $\norm{A-BL}<1$. $(A-BL)$ being Schur doesn't automatically satisfy the spectral norm criterion. Therefore, it is possible to design a different methodology to find a more suitable candidate for the gain matrix $L$, which may not minimize the LQR cost functional, but will satisfy the spectral norm criterion always. This will be investigated by the authors of this manuscript in near future. 
Moreover, it is evident that Theorem \ref{theorem4} can be immediately derived from Theorem \ref{vector_bernoulli_probability} as the spectral norm of the open loop state transition matrix $A$ is reduced to the absolute open loop gain $|\alpha|$ of the scalar plant, and the spectral norm of the closed loop state transition matrix $(A-BL)$ is reduced to the absolute closed loop gain $|\alpha-bl|$ of the scalar plant.\\
If the intermittence probability of the Bernoulli switch is below the threshold given in Theorem \ref{theorem7}, the asymptotic state covariance matrix $(P_\infty = \lim_{n\rightarrow\infty}P_n)$ can be computed by solving the following $\frac{n(n+1)}{2}$ linear equations. 
\begin{eqnarray}
 \sum_{l=2}^{n}\sum_{k=1}^{l-1}\left[p(a_{il}a_{jk}+a_{ik}a_{jl})+(1-p)(g_{il}g_{jk}+g_{ik}g_{jl})\right]s_{lk}
+\sum_{l=1}^n \left[pa_{il}a_{jl}+(1-p)g_{il}g_{jl}\right]s_{ll}+w_{ij} = s_{ij}
\end{eqnarray}
$\forall i =1,2,\dots,n$ and $j = 1,2,\dots, i$. 
Here, $a_{ij}, g_{ij}, w_{ij}, s_{ij}$ are the $(i,j)^{th}$ elements of the matrices $A, (A-BL), W, P_\infty$ respectively. These equations are obtained by expanding the generalized Lyapunov equation 
\begin{eqnarray}
pAP_\infty A'+(1-p)(A-BL)P_\infty(A-BL)'+W = P_\infty    
\end{eqnarray}
and the non-redundant equations are kept. \\
%\end{proof}
Once the expected system dynamics, and the condition of boundedness of the state covariance matrix $(P_n)$ are obtained, the result towards the design of the recursive quantizer be stated such that the system can be $\mathcal{L}_2$ stabilized even in the presence of finite capacity channel. 
However, for vector plant with input dimensionality being less than the state dimensionality $(m<n)$, it is not possible to find out the covariance matrix of the quantized signal analytically using the posterior distribution of the quantized signal following the technique of Lemma \ref{lemma2}. This is stated in the next lemma. 
\begin{lemma}
\label{lemma8}
 Lemma \ref{lemma2} cannot be extended to the vector plant with $m<n$. 
\end{lemma}
{\bf Proof}:  The proof is provided in section \ref{proof8}.

From the scalar example, it has been seen that the upper bound of the intermittence probability is an increasing function of channel capacity $\mathcal{C}$.
%********** RKL insert figure of the curve ****************
However, for vector plant with $m<n$, the upper bound of $p$ is obtained only for $\mathcal{C}\rightarrow\infty$. Clearly Theorem \ref{theorem7} is only a necessity and not sufficiency here. 
In this situation, the recursive quantization methodology to mitigate the finite capacity channel constraint can only be designed based on the dynamics of the conditional state covariance matrix when $\mathcal{C}\rightarrow\infty$. The recursion is given in Theorem \ref{theorem9}. 

\begin{theorem}
\label{theorem9}
    
    For the vector plant described in section \ref{vect_bern_sys}, for $\mathcal{C}< \infty$, the recursive update equation of the uniform quantizer step-size per dimension is given by
    \begin{eqnarray}
        {\pmb\Delta}_{k+1|{\pmb \gamma}_{0:k}}^2 = {\pmb\Delta}_{k|{\pmb \gamma}_{0:k-1}}^2 \left[{\pmb\gamma}_k |A|^{\frac{2}{n}}+(1-{\pmb\gamma}_k)|A-BL|^{\frac{2}{n}}\right]
    + 2\pi e \left(2^{-\frac{2\mathcal{C}}{n}}\right)|W|^{\frac{1}{n}}+\epsilon
    \end{eqnarray}
    Where, ${\pmb \Delta}_{k|{\pmb \gamma}_{0:k-1}}$ is the conditional quantization step-size per state dimension at $k^{th}$ time step, and $\epsilon >0$ is a small constant. 
\end{theorem}
{\bf Proof}:  The proof is provided in section \ref{proof9}.

The algorithm for this recursive quantization scheme is given in section \ref{algo}.

\subsection{Markov intermittence model}

\subsubsection{Scalar Plant}
\label{markov_scalar}
A system is taken similar to the one described for the Bernoulli intermittence model. The only difference is that the intermittent switch state ${\pmb \gamma}_k$ is considered to be a Markov chain whose transition probability matrix (TPM) is given by: 
$T = \begin{bmatrix}
    1-p & p \\
    q & 1-q \\
    \end{bmatrix} \\\\ $
, where $p,q\in[0,1]$. $p$ and $q$ need not be equal. 
The state of the Markov chain is defined by $\zeta_k=[(1-\pi_k),\pi_k]$, where $\pi_k=P({\pmb \gamma}_k=1)$.
The initial condition of the Markov chain is assumed to be $\zeta_0 =[(1-\pi_0), \pi_0]$, $\pi_0\in[0,1]$.
The following lemmas and theorems are used to obtain the conditions for ${\mathcal{L}_2} $ stabilization of the system.
\begin{lemma}
\label{lemma10}
    For the scalar plant described in section \ref{markov_scalar}, in absence of quantization (i.e., assuming $\mathcal{C} \rightarrow \infty$), the conditional state variance ${\pmb \sigma}^2_{k+1|{\pmb \gamma}_{0:k}}$ at the $(k+1)^{th}$ time step is given by:
\begin{eqnarray}
\label{scal_markov_cond_var_eq}
        {\pmb \sigma}_{k+1|{\pmb \gamma}_{0:k}}^2= \left(\prod_{j=0}^{k}{\pmb\xi}_{j}^2\right)\sigma_{0}^2 + \left[\sum_{j=0}^{k-1}\left(\prod_{i=j+1}^{k}{\pmb\xi}_{i}^2\right) + 1\right]\sigma_{\bf w}^2
    \end{eqnarray}
    , and the (expected) state variance $\sigma_{k+1}^2$ at the $(k+1)^{th}$ time step is given by:
    \begin{eqnarray}
    \label{scal_markov_var_eq}
        {\sigma}_{k+1}^2 = \left(\prod_{j=0}^{k}\omega_{j}^2\right)\sigma_{0}^2 + \left[\sum_{j=0}^{k-1}\left(\prod_{i=j+1}^{k}\omega_{i}^2\right) + 1\right]\sigma_{\bf w}^2
    \end{eqnarray}
    where, ${\pmb \xi}_{i}^2 = \alpha^2 {\pmb\gamma}_i + (\alpha-bl)^2\left(1 - {\pmb\gamma}_i\right)$, and $\omega_{i}^2 = \alpha^2\pi_i + (\alpha-bl)^2\left(1 - \pi_i\right)$.
\end{lemma}

{\bf Proof}:  The proof is provided in section \ref{proof10}.

\begin{lemma}
\label{lemma11}
For the system discussed in section \ref{markov_scalar}, if $\mathcal{C}<\infty$, and if the quantization step size at the $k^{th}$ time step is ${\pmb\Delta}_{k}$, conditioning on the intermittence random sequence, the one step conditional state variance update equation at the $(k+1)^{th}$ time step is given by: 
\begin{eqnarray}
\label{cond_var_markov_scalar_eqn}
        {\pmb\sigma}_{k+1|{\pmb\gamma}_{0:k}}^2={\pmb\xi}_k^2{\pmb\sigma}_{k|{\pmb\gamma}_{0:k-1}}^2+b^2 l^2(1-{\pmb\gamma}_k)\frac{{\pmb\Delta}_{k|{\pmb\gamma}_{0:k-1}}^2}{12}+\sigma_{\bf w}^2
    \end{eqnarray}
, and the one step (expected) state variance update equation at the $(k+1)^{th}$ time step is given by:
\begin{eqnarray}
    \label{var_markov_scal_eqn}
        \sigma_{k+1}^2 = \omega_k^2\sigma_{k}^2 + b^2l^2(1-\pi_k)\frac{\Delta_{k}^2}{12} + \sigma_{\bf w}^2
    \end{eqnarray}
\end{lemma}

{\bf Proof}:  The proof is provided in section \ref{proof11}.

\begin{lemma}
\label{lemma12}
For the system discussed in section \ref{markov_scalar}, the one step conditional quantization step size update equation for the $(k+1)^{th}$ time step is given by:

\begin{eqnarray}
\label{scal_markov_cond_quant_eq}
       {\pmb \Delta}_{k+1|{\pmb \gamma}_{0:k}}^2 =
       {\pmb \Delta}_{k|{\pmb \gamma}_{0:k-1}}^2\left[{\pmb \xi}_k^2
       +\eta G(1-{\pmb\gamma}_k)\right] 
       + \eta\sigma_{\bf w}^2 + \epsilon(1 - {\pmb \xi}_k^2)
    \end{eqnarray}

, and the one step (expected) quantization step size update equation for the $(k+1)^{th}$ time step is given by:
    \begin{eqnarray}
    \label{scal_markov_quant_eq}
        \Delta_{k+1}^2 = \Delta_{k}^2\left[\omega_k^2 + \eta G(1-\pi_k)\right] + \eta\sigma_{\bf w}^2 + \epsilon(1 - \omega_k^2)
    \end{eqnarray}.
    
    Here, $\eta = \frac{2\pi e} {2^{2\mathcal{C}}}$, $G = \frac{b^2 l^2}{12}$, and $\epsilon>0$ is a small constant. 
\end{lemma}

{\bf Proof}:  The proof is provided in section \ref{proof12}.

Based on these lemmas, the theorem on the conditions for $\mathcal{L}_2$ stabilization can be stated. 
\begin{theorem}
\label{theorem13}
   For the system discussed in section \ref{markov_scalar}, the necessary and sufficient conditions for $\mathcal{L}_2$ stabilization are given below. 
    \begin{enumerate}
        \item For $\mathcal{C} \rightarrow \infty$, the necessary and sufficient condition for $\mathcal{L}_2$ stabilization of the system is given by:
        \begin{eqnarray}
        \label{scal_markov_iff_inf}
            \frac{p}{q} \le \frac{1 - (\alpha-bl)^2}{\alpha^2 - 1}
        \end{eqnarray}
        , and if the condition is satisfied, the state variance asymptotically converges as follows:
        \begin{eqnarray}
        \label{scal_markov_iff_inf_var}
\lim_{k\rightarrow\infty}\sigma_k^2=\frac{\left(1+\frac{p}{q}\right)}{1 - (\alpha-bl)^2 - \frac{p}{q}(\alpha^2 - 1)}\sigma_{\bf w}^2
        \end{eqnarray}
        
        \item For $\frac{1}{2}\log(\frac{\alpha^2}{6\pi e})<\mathcal{C} < \infty$, the necessary and sufficient condition for $\mathcal{L}_2$ stabilization of the system is given by:
        \begin{eqnarray}
        \label{scal_markov_iff_fin}
        \frac{p}{q} < \frac{1 -(\alpha-bl)^2 - \eta G}{\alpha^2 - 1}
        \end{eqnarray}
        , and if the condition is satisfied, the state variance asymptotically converges as follows:
        \begin{eqnarray}
            \label{scal_markov_iff_fin_var}
           \lim_{k\rightarrow\infty}\sigma_k^2 = \frac{\sigma_{\bf w}^2\left(1+\frac{p}{q}\right) + \epsilon G}{\left(1+\frac{p}{q}\right)-\left[\alpha^2\frac{p}{q}+(\alpha-bl)^2+\eta G\right]}
        \end{eqnarray}
    \item If $\mathcal{C}\le \frac{1}{2}\log(\frac{\alpha^2}{6\pi e})$, the system will always be $\mathcal{L}_2$ unstable.
    \end{enumerate}
    Here, $\eta = \frac{2\pi e}{2^{2\mathcal{C}}}$, $G = \frac{b^2 l^2}{12}$, and $\epsilon>0$ is a small constant. 
\end{theorem}

{\bf Proof}:  The proof is provided in section \ref{proof13}

\subsubsection{Vector Plant}
\label{vector_markov}
The vector plant with Markov intermittence switching is given by:
\begin{eqnarray*}
    {\bf x}_{k+1} = \left\{{\pmb\gamma}_k A +(1-{\pmb \gamma}_k)(A-BL)\right\} {\bf x}_k+{\bf w}_k
\end{eqnarray*}
Here, ${\pmb\gamma}_k$ is the intermittent switch binary random variable with Markovian property as discussed in section \ref{markov_scalar}. The following lemmas will lead to the main theorems. 
\begin{lemma}
\label{lemma14}
\label{vector_markov_covariance}
For the vector plant described in section \ref{vector_markov}, the conditional state covariance matrix in the absence of quantization, (i.e, when the channel capacity $(\mathcal{C})$ is infinite) is given by:
\begin{eqnarray}
\label{vec_markov_cond_cov_eq}
   \nonumber {\bf P}_{k+1|{\pmb\gamma}_{0:k}}&=&\mathbb{E}\left({\bf x}_{k+1}{\bf x}'_{k+1}|{\pmb \gamma}_{0:k}\right)\\
    &=&\left(\prod_{j=k}^{0}{\pmb \Lambda}_j\right) P_0\left(\prod_{j=0}^{k}{\pmb\Lambda}'_j\right)
    +\left[\sum_{j=0}^{k-1}\left\{\left(\prod_{i=k}^{j+1}{\pmb\Lambda}_i\right)W\left(\prod_{i=j+1}^{k}{\pmb\Lambda}'_i\right)\right\}+W\right]
\end{eqnarray}
Where, ${\pmb\Lambda}_k = {\pmb\gamma}_k A + \left(1-{\pmb\gamma}_k\right)\left(A-BL\right)$, $P_0 = \mathbb{E}({\bf x}_0{\bf x}'_0)$, and $W = \mathbb{E}({\bf w}_k {\bf w}'_k)$. 
\end{lemma}
{\bf Proof}:  The proof is provided in section \ref{proof14}.

\begin{lemma}
\label{lemma15}
\label{vector_markov_covariance_expected}
For the vector plant described in section \ref{vector_markov} , the one step update equation of the state covariance matrix ($P_k$) in absence of quantization, (i.e, $\mathcal{C}\rightarrow\infty$) is given by:
\begin{eqnarray}
\label{vec_markov_cov_eq}
P_{k+1} =
\pi_k AP_{k}A'+(1-\pi_k)(A-BL)P_{k}(A-BL)'+W
\end{eqnarray}
where, $P_k = \mathbb{E}\left({\bf x}_k {\bf x}'_k\right)$, and $\pi_k = P({\pmb \gamma}_k =1)$. \\
\end{lemma}
{\bf Proof}:  The proof is provided in section \ref{proof15}.

Now the main theorem providing the necessary and sufficient condition for $\mathcal{L}_2$ stabilization of the vector plant without quantization under Markov intermittence is stated. 
\begin{theorem}
\label{theorem16}
\label{vector_markov_probability}
   For the system described in section \ref{vector_markov}, if $\mathcal{C}\rightarrow\infty$, the necessary and sufficient condition for the asymptotic state covariance matrix  to be bounded, and unique is given by:
\begin{eqnarray}
\label{vec_markov_iff_l2}
 \frac{p}{q}< \frac{1-\norm{A-BL}_2^2}{\norm{A}_2^2-1}
 \end{eqnarray}
 Where, $\norm{A}_2$ is the spectral/operator norm of the matrix $A$.
\end{theorem}

{\bf Proof}:  The proof is provided in section \ref{proof16}.

Moreover, it is evident that the first part of Theorem \ref{theorem13} (Eqn. \ref{scal_markov_iff_inf}) can be immediately derived from Theorem \ref{vector_markov_probability} as the spectral norm of the open loop state transition matrix $A$ is reduced to the absolute open loop gain $|\alpha|$ of the scalar plant, and the spectral norm of the closed loop state transition matrix $(A-BL)$ is reduced to the absolute closed loop gain $|\alpha-bl|$ of the scalar plant.\\ 
  If the intermittence parameters of the Markov switch ($p$ and $q$) satisfy the necessary and sufficient condition as given in Theorem \ref{theorem16}, the asymptotic state covariance matrix $(P_\infty)$ can be computed by solving the following $\frac{n(n+1)}{2}$ linear equations. 
\begin{eqnarray}
\sum_{l=2}^{n}\sum_{k=1}^{l-1}\left[p(a_{il}a_{jk}+a_{ik}a_{jl})+q(g_{il}g_{jk}+g_{ik}g_{jl})\right]s_{lk}+
\sum_{l=1}^n \left[pa_{il}a_{jl}+qg_{il}g_{jl}\right]s_{ll} = (p+q)(s_{ij}-w_{ij})
\end{eqnarray}
$\forall i =1,2,\dots,n$ and $j = 1,2,\dots, i$. 
Here, $a_{ij}, g_{ij}, w_{ij}, s_{ij}$ are the $(i,j)^{th}$ elements of the matrices $A, (A-BL), W,$ and $P_\infty$ respectively. These equations are obtained by expanding the following generalized asymptotic Lyapunov equation and keeping the non-redundant terms.
\begin{eqnarray*}
\pi_\infty AP_\infty A'+(1-\pi_\infty)(A-BL)P_\infty(A-BL)'+W = P_\infty \end{eqnarray*} 

Once the system dynamics, and the conditions for convergence of the state covariance matrix are obtained, the main result towards the design of the recursive quantizer be stated.
However, for vector plant with $m<n$, it is not possible to develop the covariance matrix of the quantized signal using the posterior distribution of the quantized signal. 
\begin{lemma}
\label{lemma17}
 Lemma \ref{lemma11} cannot be extended to the vector plant with $m<n$. 
\end{lemma}
{\bf Proof}:  The proof is provided in section \ref{proof17}.

In this situation, the quantization to mitigate the finite capacity channel constraint can only be designed based on the dynamics of the conditional state covariance matrix when the channel is of infinite capacity. The result is given in the theorem below. 
\begin{theorem}
\label{theorem18}
    
    For the system discussed in section \ref{vector_markov}, for $\mathcal{C}< \infty$, the recursive update equation of the uniform quantizer step-size per dimension is given by
    \begin{eqnarray}
        {\pmb\Delta}_{k+1|{ {\pmb\gamma}_{0:k}}}^2 = {\pmb\Delta}_{k|{{\pmb\gamma}_{0:k-1}}}^2 \left[{\pmb\gamma}_k |A|^{\frac{2}{n}}+(1-{\pmb\gamma}_k)|A-BL|^{\frac{2}{n}}\right]
        + 2\pi e \left(2^{-\frac{2\mathcal{C}}{n}}\right)|W|^{\frac{1}{n}}+\epsilon
    \end{eqnarray}
    Where, ${\pmb\Delta}_{k| {\pmb\gamma}_{0:k-1}}$ is the conditional quantization step-size per state dimension at the $k^{th}$ time step, and $\epsilon >0$ is a small constant. 
\end{theorem}
{\bf Proof}:  The proof is provided in section \ref{proof18}.

Clearly for $\mathcal{C}<\infty$ scenario, the condition of $\mathcal{L}_2$ stabilization of the vector plant under Markov intermittence as stated in Theorem \ref{theorem16} is only necessary, and not sufficient. The algorithm for this recursive quantization scheme is given in section \ref{algo}.
\subsection{Algorithms}
\label{algo}
Since the recursive quantization procedures are same for Bernoulli and Markov intermittence models, the algorithm is presented for both of them together. The algorithm for the scalar plant is presented in Algorithm \ref{algo:scal}, and the algorithm for the vector plant is presented in Algorithm \ref{algo:vec}.
\begin{algorithm}[H]
\caption{Recursive quantization for the scalar plant}\label{algo:scal}
\begin{algorithmic}
\STATE INPUT: $\alpha, b, \mathcal{C}$ (system parameters)
\STATE INPUT: $x_0, \sigma_0^2, \sigma_{\bf w}^2$ (Initial condition and noise)
\STATE INPUT: $p$ (for Bernoulli switch)\\
\hspace{1.2cm}$p,q$ (for Markov switch)\\
INPUT: $N$ (Stopping time)
\STATE INITIALIZE: $k=0$, $\Delta_0^2 = \eta \sigma_0^2+\epsilon$ \\
(Here, $\eta = \frac{2\pi e}{2^{2\mathcal{C}}}$, $\epsilon>0$)
\STATE WHILE $k<N$
\STATE SWITCH: Draw $\gamma_k$ (Bernoulli or Markov)
\STATE IF $\gamma_k = 0$
\STATE QUANTIZATION: $x_k^{\Delta}=Q_{\text {uniform}}\left(x_k,\Delta_{k|{\gamma}_{0:k-1}}\right)$
\STATE ENCODING-DECODING: Any standard protocol to send the signal through the channel
\STATE CONTROL: $u_k = -lx_k^{\Delta}$\\
($l$ is the infinite horizon LQR control gain.)
\STATE PLANT: $x_{k+1}=\alpha x_k+ b u_k + w_k$
\STATE ELSE IF
\STATE PLANT: $x_{k+1} = \alpha x_k+w_k$
\STATE END IF
\STATE QUANTIZATION STEP SIZE: \\
    $\Delta_{k+1|{\gamma}_{0:k}}^2 = \Delta_{k|{\gamma}_{0:k-1}}^2 \left[\alpha^2\gamma_k+\left((\alpha-bl)^2+\eta \frac{b^2l^2}{12}\right)(1-\gamma_k)\right]+\eta \sigma_{\bf w}^2+\epsilon\left[1-\alpha^2\gamma_k-(\alpha-bl)^2(1-\gamma_k)\right]$
\STATE RETURN: $x_{k+1}, \Delta_{k+1|{\gamma}_{0:k}}$
\STATE END WHILE
\STATE ESTIMATE: $\sigma_N^2$
\end{algorithmic}
\end{algorithm}
\begin{algorithm}[H]
\caption{Recursive quantization for the vector plant}\label{algo:vec}
\begin{algorithmic}
\STATE INPUT: $A, B, n, \mathcal{C}$ (system parameters)
\STATE INPUT: $x_0, P_0, W$ (Initial condition and noise)
\STATE INPUT: $p$ (for Bernoulli switch)\\
\hspace{1.2cm}$p,q$ (for Markov switch)\\
INPUT: $N$ (Stopping time)
\STATE INITIALIZE: $k=0$, $\Delta_0^2 = \eta |P_0|^{\frac{1}{n}}+\epsilon$ \\
(Here, $\eta = \frac{2\pi e}{2^{2\mathcal{C}}}$, $\epsilon>0$)
\STATE WHILE $k<N$
\STATE SWITCH: Draw $\gamma_k$ (Bernoulli or Markov)
\STATE IF $\gamma_k = 0$
\STATE QUANTIZATION: $x_k^{\Delta}=Q_{\text {uniform}}\left(x_k,\Delta_{k|{\gamma}_{0:k-1}}\right)$
\STATE ENCODING-DECODING: Any standard protocol to send the signal through the channel
\STATE CONTROL: $u_k = -Lx_k^{\Delta}$\\
($L$ is the infinite horizon LQR control gain matrix.)
\STATE PLANT: $x_{k+1}=Ax_k+ B u_k + w_k$
\STATE ELSE IF
\STATE PLANT: $x_{k+1} = A x_k+w_k$
\STATE END IF
\STATE QUANTIZATION STEP SIZE PER DIMENSION: \\
    $\Delta_{k+1|{\gamma}_{0:k}}^2 = \Delta_{k|\gamma_{0:k-1}}^2 \left[\gamma_k|A|^{\frac{2}{n}}+(1-\gamma_k)|A-BL|^{\frac{2}{n}}\right] +2\pi e (2^{-\frac{2\mathcal{C}}{n}})|W|^{\frac{1}{n}}+\epsilon$
\STATE RETURN: $x_{k+1}, \Delta_{k+1|\gamma_{0:k}}$
\STATE END WHILE
\STATE ESTIMATE: $P_N$

\end{algorithmic}
%\label{alg1}
\end{algorithm}

 In the algorithms the symbols are not bold. This is to signify that here the realized values are mentioned, not the random variables.

\section{Examples}
\label{simulation}
\subsection{Bernoulli intermittence}
\begin{figure}[H]
\centering
\includegraphics[width=3.4in]{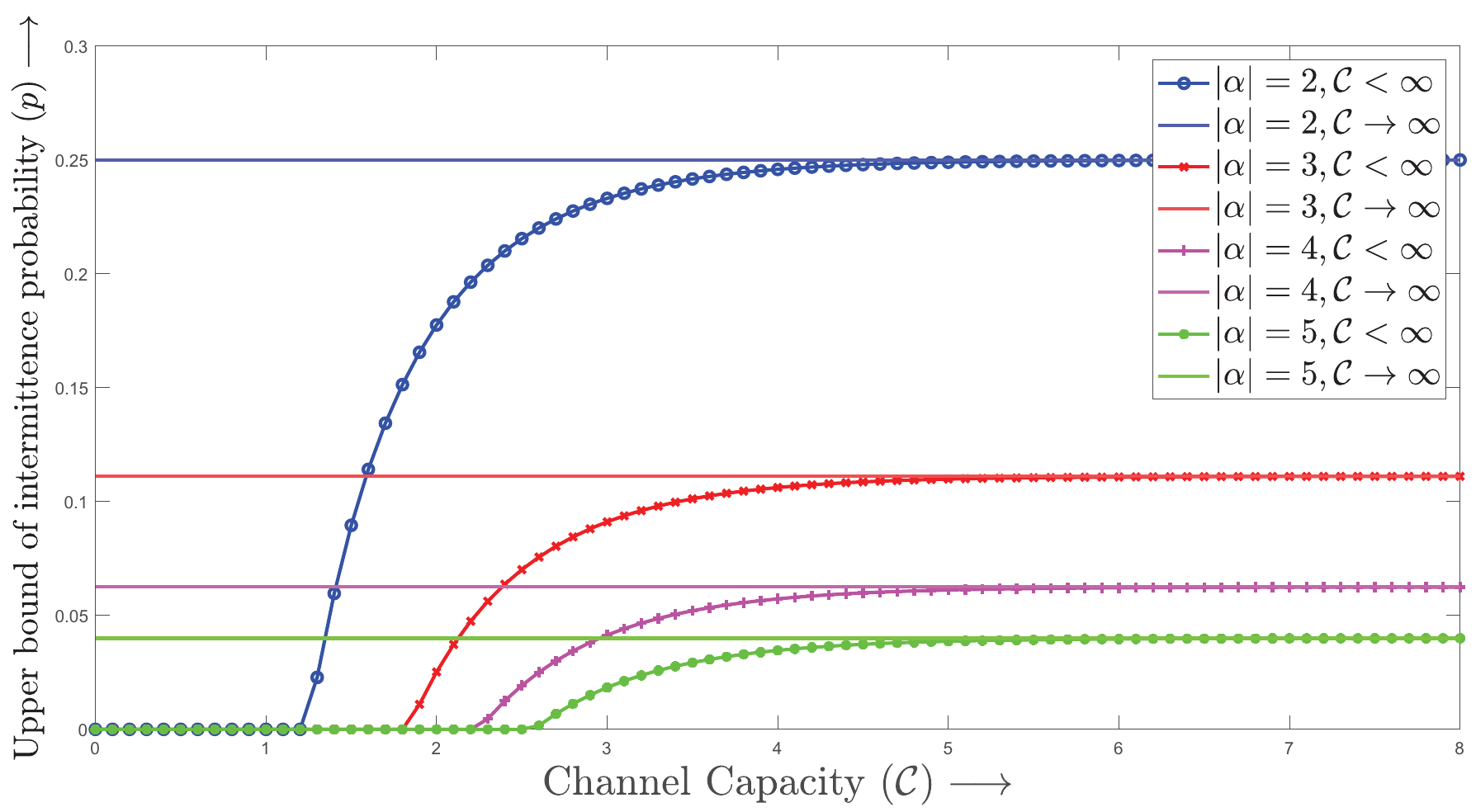}
\caption{Upper bounds of $p$ for $\mathcal{L}_2$ stability of the system under Bernoulli intermittence}
\label{p_C_bern}
\end{figure}

To study the scalar plant with Bernoulli intermittence switching, first it is pertinent to study Theorem \ref{theorem4}. For the entire range of the channel capacity i.e., $\mathcal{C} \in [0,\infty)$, the upper bound of the intermittence parameter $p$ is plotted in Fig. \ref{p_C_bern} for different absolute values of the plant parameter $|\alpha|$. All the curves are reaching to the corresponding asymptotes.

\begin{figure}[H]
\centering
\includegraphics[width=3.4in]{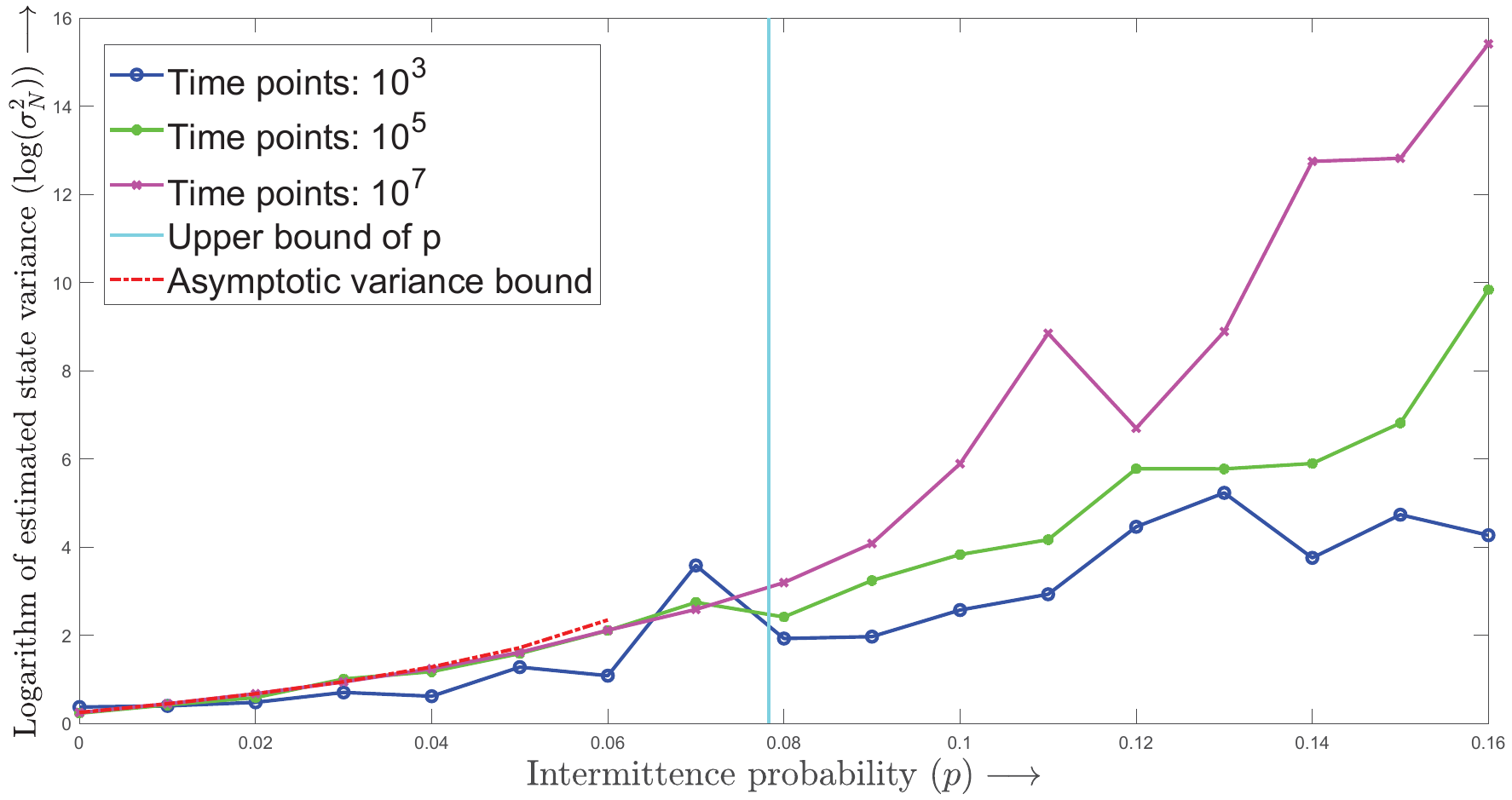}
\caption{State variance simulation and the theoretical asymptotic bound for scalar plant under Bernoulli intermittence}
\label{LQG1}
\end{figure}

\begin{figure}[H]
\centering
\includegraphics[width=3.5in]{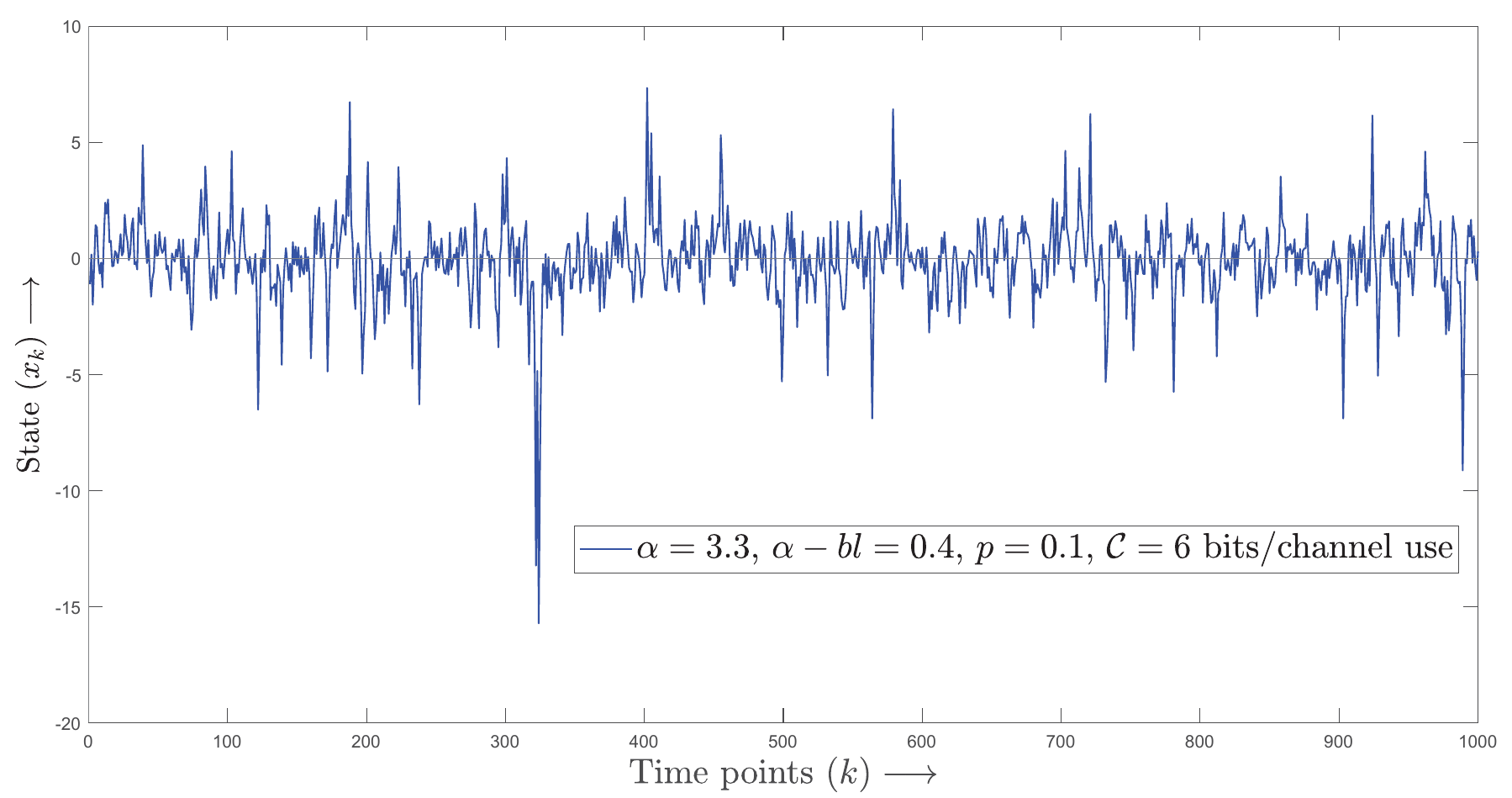}
\caption{Stable loop state trajectory under Bernoulli intermittence}
\label{dynamics1}
\end{figure}

\begin{figure}[H]
\centering
\includegraphics[width=3.5in]{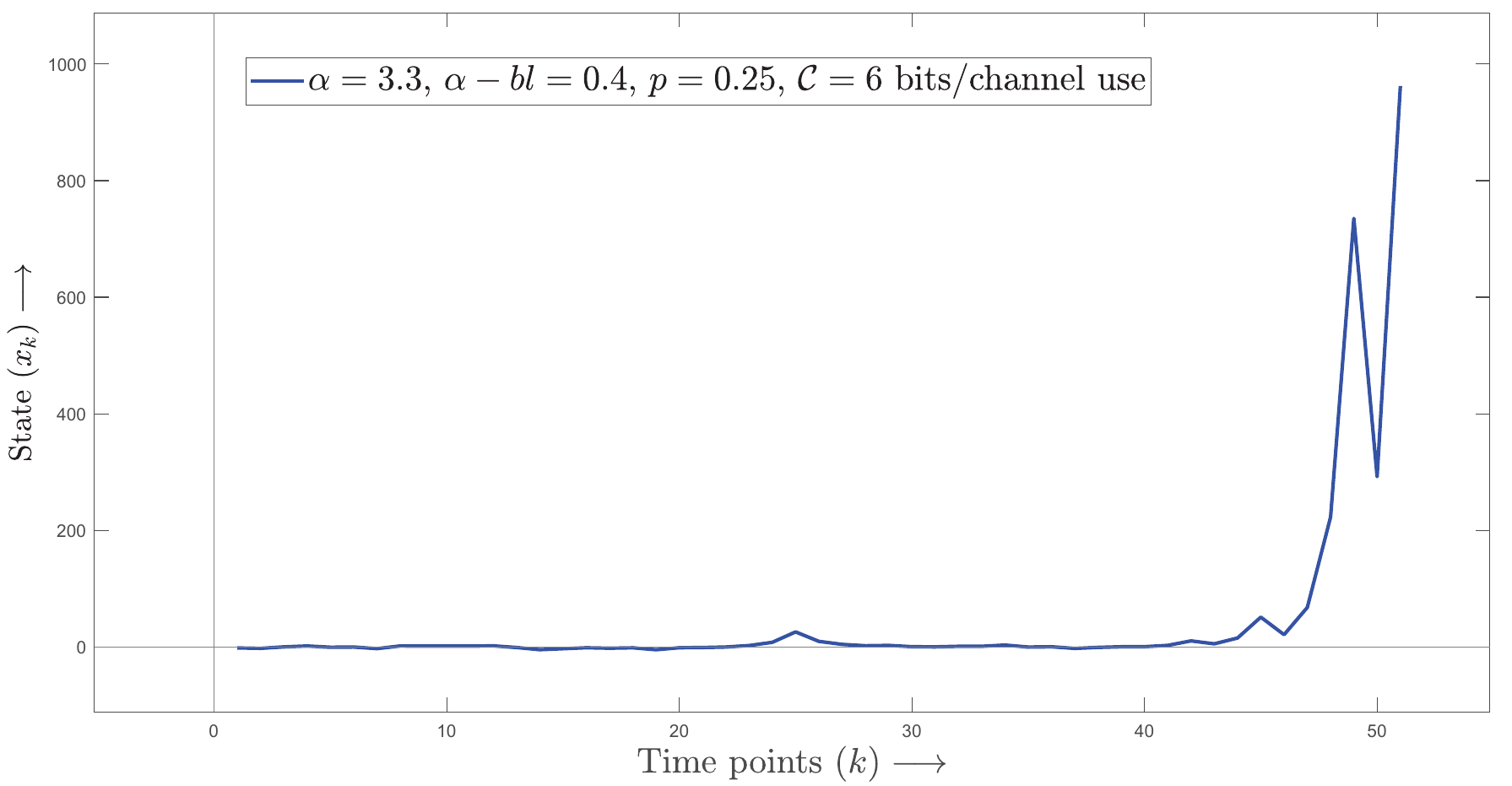}
\caption{Unstable loop state trajectory under Bernoulli intermittence}
\label{dynamics2}
\end{figure}

The next interesting study is of the convergences and divergences of sample state variances for different choices of $p$. When $p$ is below the upper bound of $p$ as found out from Theorem \ref{theorem4}, and shown in Fig. \ref{p_C_bern}, the sample variances for different time samples converge to the theoretically predicted value. If $p$ goes above the bound, the sample state variance starts diverging for larger time points. This result is shown in Fig. \ref{LQG1}. In this experiment the following values are used for the system: open loop gain $\alpha = 3.3$, closed loop gain $(\alpha-bl)=0.4$, $\sigma_0^2 = 4$, and $\sigma_{\bf w}^2 = 1$. 
Although Fig. \ref{p_C_bern} and Fig. \ref{LQG1} demonstrate the asymptotic behaviors and the bounds, it is still important to show the dynamics of the system, albeit for just two sample trajectories of the stochastic system. 
First, a convergent trajectory is shown by choosing the parameters in the convergent region.  For $\alpha = 3.3, (\alpha-bl)=0.4, p=0.1, \mathcal{C} = 6$ bits/channel use, the state trajectory sample function is not diverging for $1000$ time points as seen in Fig. \ref{dynamics1}.
However, by modifying intermittence probability to $p=0.3$, the divergence of the sample function is observed at a time point of $51$ as seen in Fig. \ref{dynamics2}. Although these simulations are only for tutorial purpose. A few sample functions don't say much about stochastic dynamics. 

\subsection{Markov intermittence}
For the scalar plant with Markov intermittent switching, Theorem \ref{theorem13} is studied. For the entire range of the channel capacity i.e., $\mathcal{C} \in [0,\infty)$, the upper bound of ($\frac{p}{q}$) is plotted for different absolute values of the plant parameter $|\alpha|$. All the curves are reaching to the corresponding asymptotes.

\begin{figure}[H]
\centering
\includegraphics[width=3.4in]{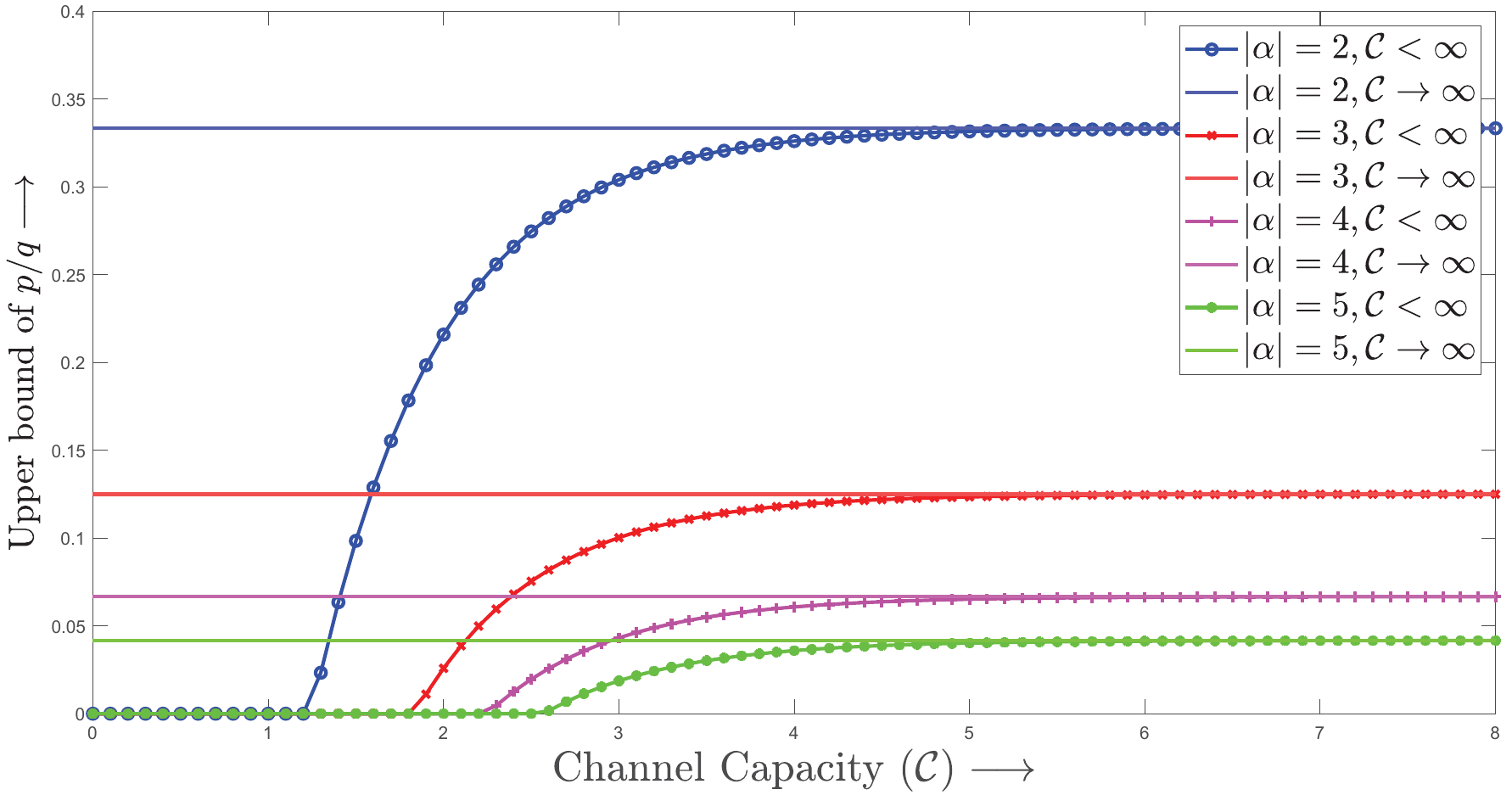}
\caption{Upper bounds of $(\frac{p}{q})$ for $\mathcal{L}_2$ stability of the system under Markov intermittence}
\label{p_C_markov}
\end{figure}

\begin{figure}[H]
\centering
\includegraphics[width=3.5in]{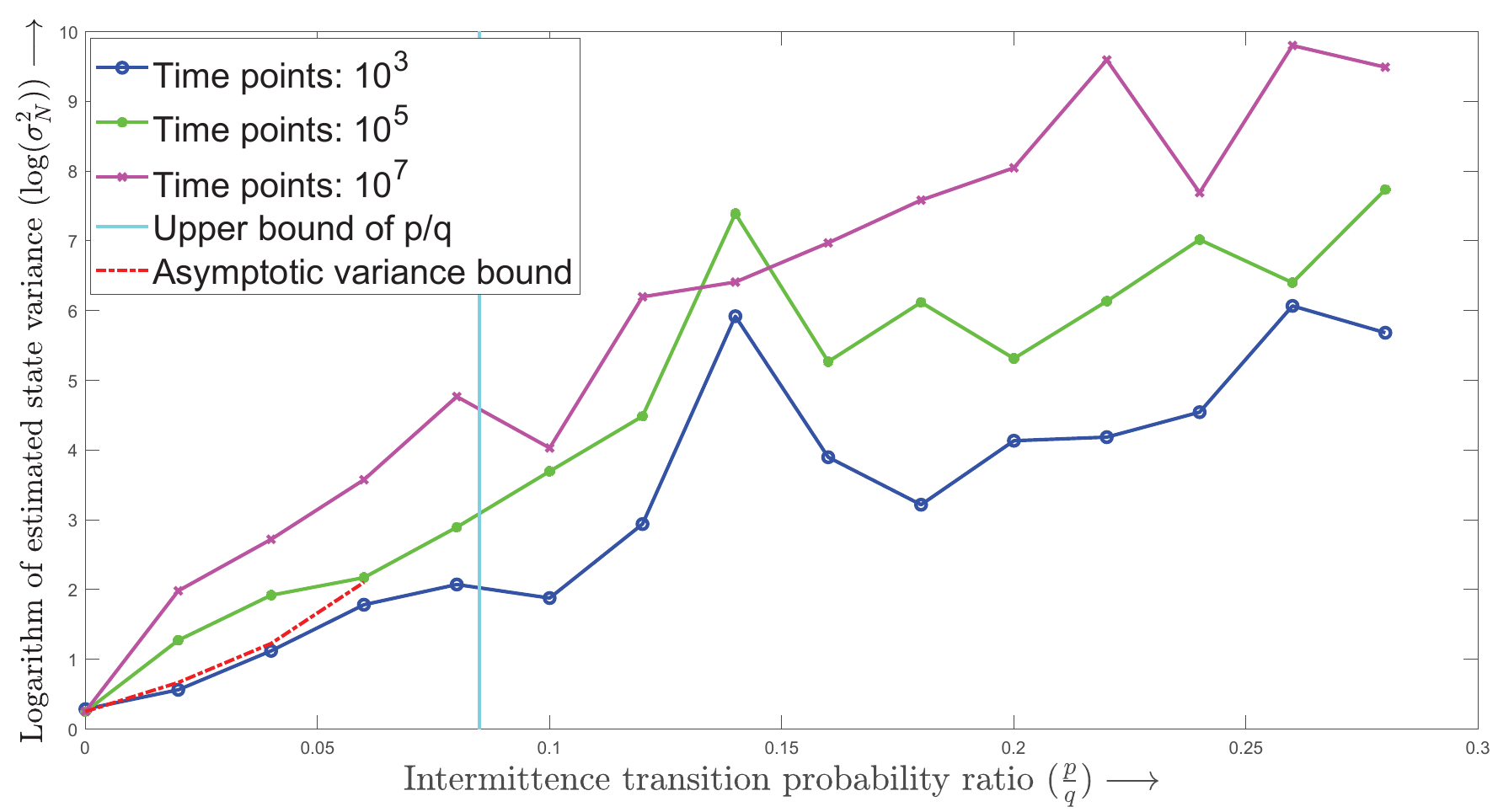}
\caption{State variance simulation and the theoretical asymptotic bound for scalar plant under Markov intermittence}
\label{LQG2}
\end{figure}

The next simulation study is of convergences and divergences of sample state variances for different choices of $(\frac{p}{q})$. When $\frac{p}{q}$ is below the upper bound of $\frac{p}{q}$ as found out from Theorem \ref{theorem13}, and shown in Fig. \ref{p_C_markov}, the sample variances for different time samples converge to the theoretically predicted value. If $p$ goes above the bound, the sample state variance starts diverging for larger time points. This result is shown in Fig. \ref{LQG2}. In this experiment the following values are used for the system: open loop gain $\alpha = 3.3$, closed loop gain $(\alpha-bl)=0.4$, $\sigma_0^2 = 1$, and $\sigma_{\bf w}^2 = 1$. 

\begin{figure}[h]
\centering
\includegraphics[width=3.5in]{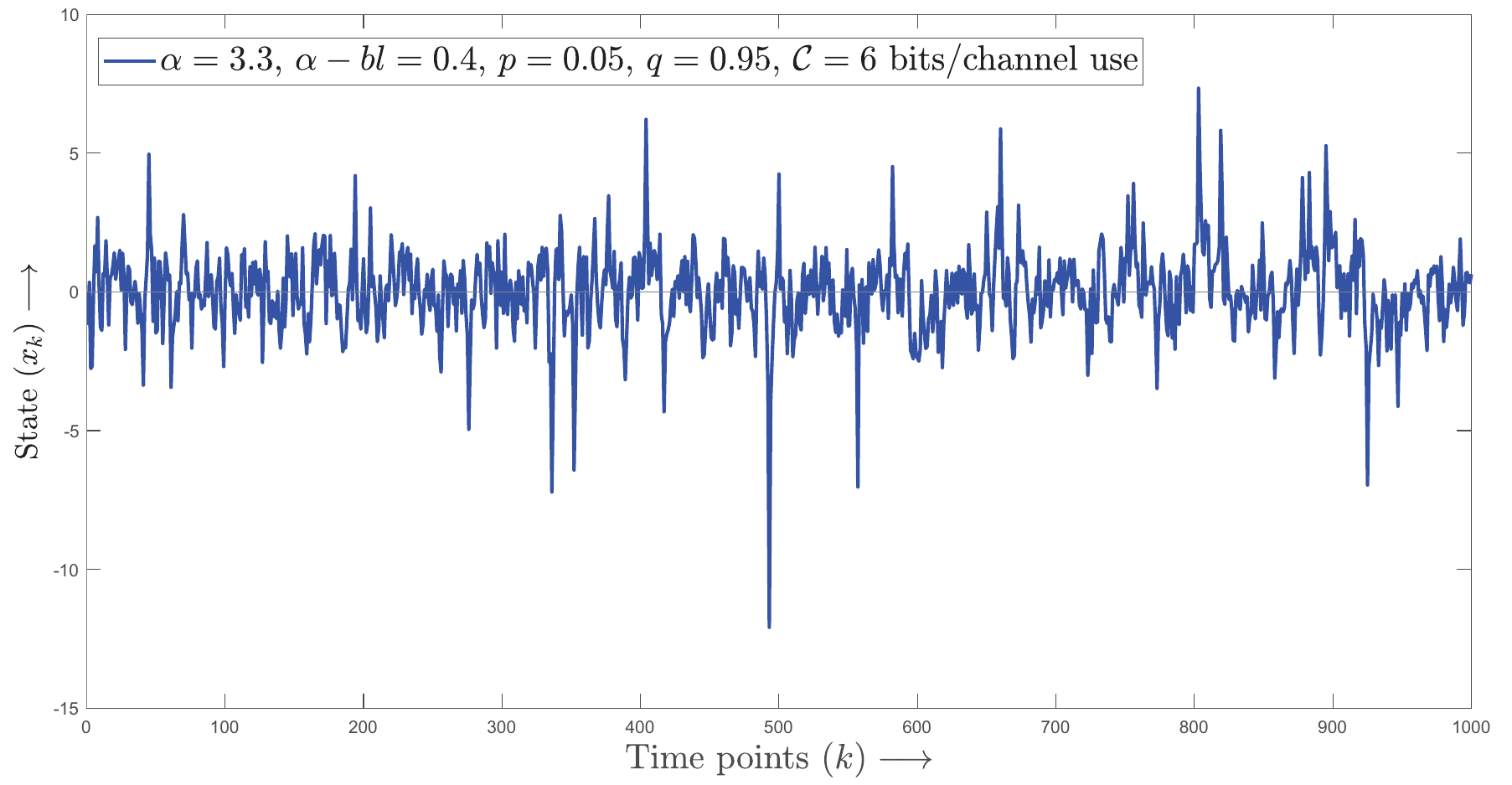}
\caption{Stable loop state trajectory under Markov intermittence}
\label{dynamics3}
\end{figure}

\begin{figure}[H]
\centering
\includegraphics[width=3.5in]{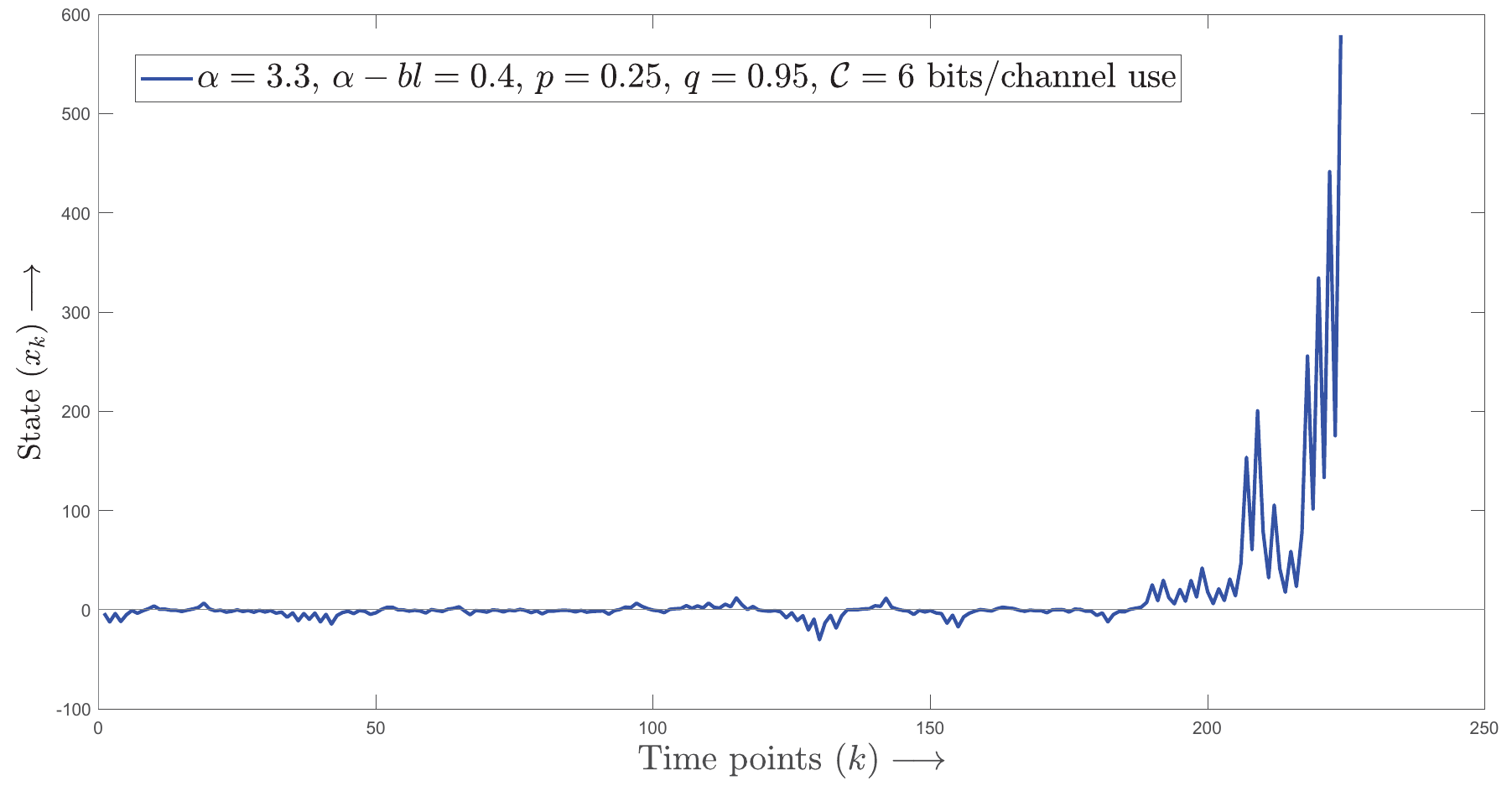}
\caption{Unstable loop state trajectory under Markov intermittence}
\label{dynamics4}
\end{figure}

First, a convergent trajectory is shown by choosing the parameters in the convergent region.  For $\alpha = 3.3, (\alpha-bl)=0.4, p=0.05, q=0.95, \mathcal{C} = 6$ bits/channel use, the state trajectory sample function is not diverging for $1000$ time points as seen in Fig. \ref{dynamics3}.
However, by modifying one parameter $p$ to $p=0.25$, the divergence of the sample function is observed at a time point of $224$ as seen in Fig. \ref{dynamics4}. 

Unfortunately, for vector system, there is no necessary and sufficient condition on the intermittence parameters for a finite capacity channel setting. So, simulations such as Fig. \ref{LQG1} or Fig. \ref{LQG2} are not possible. So, only sample function trajectory simulation could have been shown, but that would have been similar to the scalar ones. Hence, they are omitted. 

\section{Conclusion}
\label{conclusion}
The work presented in this manuscript is novel and it opens up various fundamental questions in the intersection of control and communication theory. One open question is already discussed after stating Theorem \ref{theorem7}. The design of an optimal controller for the intermittent vector system such that the spectral norm of the closed loop gain matrix remains smaller than unity is an open problem to the best of the knowledge of the authors. Moreover, the sufficient condition of intermittence parameters for the vector plant with a finite capacity channel is still open. The lower bound on capacity for the vector plant is also not figured out. Furthermore, the immediate next big problem is of output feedback. The joint behavior of the Kalman filter and the controller should be studied under the given constraints. The finite horizon counterpart of the problems are also important. On the quantization part, non-uniform entropy coded quantization needs to be undertaken when the capacity value is small. Systems with nonlinear plant, non-Gaussian additive noise, multiplicative noise, or time varying channels are much harder problems which can be investigated in future. Finally, robust optimal control strategies such as $H_\infty$ can be researched on under the constraints of intermittent observations and finite capacity channel. \\

\section{Proofs of the main results}
\label{appendix}
\subsection{Proof of Lemma \ref{lemma1}}
\label{proof1}
    
The intermittent system dynamics under $\mathcal{C}\rightarrow\infty$ assumption is described by:
    \begin{eqnarray*}
        {\bf x}_{k+1} &=&  \alpha {\pmb \gamma}_k {\bf x}_k + (\alpha-bl) (1-{\pmb \gamma}_k) {\bf x}_k + {\bf w}_k \\
        &=& \left[\alpha {\pmb \gamma}_k + (\alpha-bl) (1 - {\pmb\gamma}_k) \right] {\bf x}_k + {\bf w}_k \\
        &=& {\pmb \xi}_k {\bf x}_k + {\bf w}_k 
    \end{eqnarray*}
    where, ${\pmb \xi}_k$ is a Bernoulli random variable, and $l$ is the infinite horizon LQR gain. Now, ${\pmb \xi}_{k}^2 = \alpha^2 {\pmb \gamma}_k + (\alpha-bl)^2 (1 - {\pmb \gamma}_k)$. Clearly, ${\pmb \xi}_{k}^2$ is also a Bernoulli random variable and $\mathbb{E}\left({\pmb \xi}_{k}^2{\pmb \xi}_{j}^2\right) = 0\forall j\ne k$. Therefore  from stationarity of ${\pmb \xi}_{k}^2$, define $\omega^2:= \mathbb{E}({\pmb \xi}_{k}^2) = \alpha^2 p + (\alpha-bl)^2 (1 - p)$. 
So, the one step conditional variance update equation at the $(k+1)^{th}$ time step is given by: 
\begin{eqnarray*}
    \mathbb{E}\left({\bf x}_{k+1}^2|{\pmb\gamma}_k\right) = \left[\alpha^2 {\pmb \gamma}_k + (\alpha-bl)^2 (1 - {\pmb\gamma}_k)\right]\mathbb{E}\left({\bf x}_{k}^2\right) + \sigma_{\bf w}^2
\end{eqnarray*}
Conditioning over ${\pmb\gamma}_{0:k-1}$, the conditional state variance is computed as:
\begin{eqnarray*}
    {\pmb \sigma}_{k+1|{\pmb\gamma}_{0:k}}^2=\mathbb{E}\left[{\bf x}_{k+1}^2|{\pmb\gamma}_{0:k}\right] = {\pmb\xi}_{k}^2 \mathbb{E}\left[{\bf x}_{k}^2|{\pmb\gamma}_{0:k-1} \right] + \sigma_{\bf w}^2 \\
    = \left(\prod_{j=0}^{k} {\pmb\xi}_{j}^2 \right)\sigma_{0}^2 + \left[\sum_{j=0}^{k-1}\left( \prod_{i=j+1}^{k} {\pmb\xi}_{i}^2 \right) + 1 \right] \sigma_{\bf w}^2
\end{eqnarray*}
The state variance at $(k+1)^{th}$ time step is computed by taking expectation of the conditional state variance with respect to ${\pmb\gamma}_{0:k}$.
\begin{eqnarray*}
    \sigma_{k+1}^2 &=& \mathbb{E}\left[\mathbb{E}({\bf x}_{k+1}^2|{\pmb\gamma}_{0:k})\right] \\
    &=& \left[\omega^{2(k+1)}\sigma_{0}^2 + \left(1 + \omega^2 + \omega^4 + ... + \omega^{2k}\right)\sigma_{\bf w}^2 \right] \\
    &=& \omega^{2(k+1)}\sigma_{0}^2 + \frac{1-\omega^{2(k+1)}}{1-\omega^2}\sigma_{\bf w}^2 \\
%    &=& \omega^{2n}\sigma_{0}^2 + \frac{1 - \omega^{2n}}{1 - \omega^2}\sigma_{w}^2
\end{eqnarray*} 
That concludes the proof. \hspace{12.3cm}$\blacksquare$

\subsection{Proof of Lemma \ref{lemma2}}
\label{proof2}
     Considering an infinite horizon LQR control formulation, the random control input is ${\bf u}_k = -l {\bf x}_k$. ${\bf x}_k$ is quantized to ${\bf x}_{k}^{\Delta}$. Therefore, the quantization error is ${\bf e}_k = {\bf x}_k - {\bf x}_{k}^{\Delta}$.

The closed loop equation is given by:
\begin{eqnarray*}
    {\bf x}_{k+1} &=& \alpha {\bf x}_k + b(-l {\bf x}_{k}^{\Delta}) + {\bf w}_k \\
    &=& \alpha {\bf x}_k - bl({ \bf x}_k - {\bf e}_k) + {\bf w}_k \\
    &=& (\alpha-bl) {\bf x}_k + bl {\bf e}_k + {\bf w}_k
\end{eqnarray*}
The open loop equation is given by:
\begin{eqnarray*}
    {\bf x}_{k+1} = \alpha {\bf x}_k + {\bf w}_k
\end{eqnarray*}
Thus, the combined random dynamics can be expressed as: 
\begin{eqnarray*}
    {\bf x}_{k+1} &=& \left[\alpha {\pmb \gamma}_k + (\alpha-bl) (1 - {\pmb\gamma}_k)\right]{\bf x}_k + bl(1 - {\pmb\gamma}_k){\bf e}_k + {\bf w}_k \\ 
    &=& {\pmb\xi}_k {\bf x}_k + bl(1 - {\pmb\gamma}_k){\bf e}_k + {\bf w}_k
\end{eqnarray*}
 At the $(k+1)^{th}$ time step, the `typical' domain of the conditional probability density function $f_{{\bf x}_{k+1}|{\pmb\gamma}_{0:k}}(x_{k+1}|{\pmb \gamma}_{0:k})$ is divided uniformly into ${\bf n}_{k+1|{\pmb \gamma}_{0:k}}$ number of bins for quantization with the width of each bin being ${\pmb\Delta}_{k+1|{\pmb \gamma}_{0:k}}$. At the $(k+1)^{th}$ time step, consider the $j^{th}$ bin $\mathcal{R}_{k+1|{\pmb \gamma}_{0:k}}^j$. If ${\bf x}_{k+1} \in \mathcal{R}_{k+1|{\pmb\gamma}_{0:k}}^j$, assign the quantized value ${\bf x}_{k+1}^{\Delta} = {\bf x}_{k+1|{\pmb \gamma}_{0:k}}^j$, where ${\bf x}_{k+1|{\pmb \gamma}_{0:k}}^j$ is chosen to be the midpoint of the interval $\mathcal{R}_{k+1|{\pmb \gamma}_{0:k}}^j$. 
The conditional density of ${\bf x}_{k+1}$ given the bin $\mathcal{R}_{k+1|{\pmb\gamma}_{0:k}}^j$ is:
\begin{eqnarray}
  \nonumber  &&f_{{\bf x}_{k+1}|{\bf x}_{k+1}^{\Delta} = {\bf x}_{k+1|{\pmb\gamma}_{0:k}}^j}\left(x_{k+1}|{\bf x}_{k+1}^{\Delta} = {\bf x}_{k+1|{\pmb \gamma}_{0:k}}^j\right) \\
  \nonumber  &=&\frac{f_{{\bf x}_{k+1}|{\pmb\gamma}_{0:k}}(x_{k+1}|{\pmb\gamma}_{0:k})}{P({\bf x}_{k+1}^{\Delta} = {\bf x}_{k+1|{\pmb \gamma}_{0:k}}^j)} \\
    \nonumber&=& \frac{f_{{\bf x}_{k+1}|{\pmb\gamma}_{0:k}}(x_{k+1}|{\pmb\gamma}_{0:k})}{\int_{\mathcal{R}_{k+1|{\pmb\gamma}_{0:k}}^j}f_{{\bf x}_{k+1}|{\pmb\gamma}_{0:k}}(x_{k+1}|{\pmb\gamma}_{0:k}) dx_{k+1}} .
\end{eqnarray}
The conditional density function of the error ${\bf e}_{k+1}$ given the bin $\mathcal{R}_{k+1|{\pmb\gamma}_{0:k}}^j$ is: 
\begin{eqnarray}
  \nonumber  &&f_{{\bf e}_{k+1}|{\bf x}_{k+1}^{\Delta} = {\bf x}_{k+1|{\pmb\gamma}_{0:k}}^j}\left(e_{k+1}|{\bf x}_{k+1}^{\Delta} = {\bf x}_{k+1|{\pmb\gamma}_{0:k}}^j\right)\\
  \nonumber  &=& f_{{\bf e}_{k+1}|{\bf x}_{k+1}^{\Delta} = {\bf x}_{k+1|{\pmb\gamma}_{0:k}}^j}\left(x_{k+1} - {\bf x}_{k+1|{\pmb\gamma}_{0:k}}^{j}|{\bf x}_{k+1}^{\Delta} = {\bf x}_{k+1|{\pmb\gamma}_{0:k}}^j\right) \\
   \nonumber &=& f_{{\bf x}_{k+1}|{\bf x}_{k+1}^{\Delta} = {\bf x}_{k+1|{\pmb \gamma}_{0:k}}^j}\left(x_{k+1}|{\bf x}_{k+1}^{\Delta} = {\bf x}_{k+1|{\pmb \gamma}_{0:k}}^j\right)
   \end{eqnarray}
For large enough capacity, if ${\bf x}_{k+1} \in \mathcal{R}_{k+1|{\pmb \gamma}_{0:k}}^j$,
\begin{eqnarray*}
&&f_{{\bf x}_{k+1}|{\bf x}_{k+1}^{\Delta} = {\bf x}_{k+1|{\pmb \gamma}_{0:k}}^j}\left(x_{k+1}|{\bf x}_{k+1}^{\Delta} = {\bf x}_{k+1|{\pmb \gamma}_{0:k}}^j\right)\\
&=& f_{{\bf x}_{k+1}|{\bf x}_{k+1}^{\Delta} = {\bf x}_{k+1|{\pmb \gamma}_{0:k}}^j}\left({\bf x}_{k+1|{\pmb\gamma}_{0:k}}^j|{\bf x}_{k+1}^{\Delta} = {\bf x}_{k+1|{\pmb \gamma}_{0:k}}^j\right)
\end{eqnarray*}
. Therefore,
\begin{eqnarray*}
    f_{{\bf e}_{k+1}|{\bf x}_{k+1}^{\Delta} = {\bf x}_{k+1|{\pmb\gamma}_{0:k}}^j}\left({ e}_{k+1}|{\bf x}_{k+1}^{\Delta} = {\bf x}_{k+1|{\pmb \gamma}_{0:k}}^j\right) 
    = \frac{1}{{\pmb\Delta}_{k+1|{\pmb\gamma}_{0:k}}}
\end{eqnarray*}
Hence, the marginal density function of ${\bf e}_{k+1}$ over the choice of $j\in\{1,2,\dots,{\bf n}_{k+1|{\pmb\gamma}_{0:k}}\}$ is given by:
\begin{eqnarray*}
    &&f_{{\bf e}_{k+1|{\pmb\gamma}_{0:k}}}(e_{k+1}|{\pmb\gamma}_{0:k})\\
    &=&\sum_{j=1}^{{\bf n}_{k+1|{\pmb\gamma}_{0:k}}} \frac{1}{{\pmb\Delta}_{k+1|{\pmb\gamma}_{0:k}}}P({\bf x}_{k+1}^{\Delta} = {\bf x}_{k+1|{\pmb\gamma}_{0:k}}^j) \\
    &=& \frac{1}{{\pmb\Delta}_{k+1|{\pmb\gamma}_{0:k}}} \sum_{j=1}^{{\bf n}_{k+1|{\pmb\gamma}_{0:k}}} P({\bf x}_{k+1}^{\Delta} = {\bf x}_{k+1|{\pmb\gamma}_{0:k}}^j)\\ 
    &=& \frac{1}{{\pmb\Delta}_{k+1|{\pmb\gamma}_{0:k}}}
\end{eqnarray*}
Thus, ${\bf e}_{k+1} \sim {\mathcal{U}}\left(\left[-\frac{{\pmb\Delta}_{k+1|{\pmb\gamma}_{0:k}}}{2},+\frac{{\pmb\Delta}_{k+1|{\pmb\gamma}_{0:k}}}{2}\right]\right)$ is a (conditionally) uniform random variable with $\mathbb{E}({\bf e}_{k+1}|{\pmb\gamma}_{0:k}) = 0$, and $Var\left({\bf e}_{k+1}|{\pmb\gamma}_{0:k}\right) = \frac{{\pmb\Delta}_{k+1|{\pmb\gamma}_{0:k}}^2}{12}$. The conditional state variance update equation is: 
\begin{eqnarray*}
    {\pmb \sigma}_{k+1|{\pmb\gamma}_{0:k}}^2 = \mathbb{E}\left[{\bf x}_{k+1}^2|{\pmb\gamma}_{0:k}\right] 
    = {\pmb \xi}_{k}^2{\pmb \sigma}_{k|{\pmb\gamma}_{0:k-1}}^2 + b^2 l^2(1 - {\pmb\gamma}_k) \frac{{\pmb\Delta}_{k|{\pmb\gamma}_{0:k-1}}^2}{12} + \sigma_{\bf w}^2 
    \end{eqnarray*}
    The state variance update equation is obtained by taking expectation over ${\pmb\gamma}_{0:k}$.
    \begin{eqnarray*}
     \sigma_{k+1}^2 = \mathbb{E}\left[{\pmb\sigma}_{k+1|{\pmb\gamma}_{0:k}}^2\right] 
    = \omega^2\sigma_{k}^2 + b^2 l^2 (1 - p)\frac{\Delta_{k}^2}{12} + \sigma_{\bf w}^2
\end{eqnarray*}

\begin{figure}[h]
\centering
\includegraphics[width=3.5in]{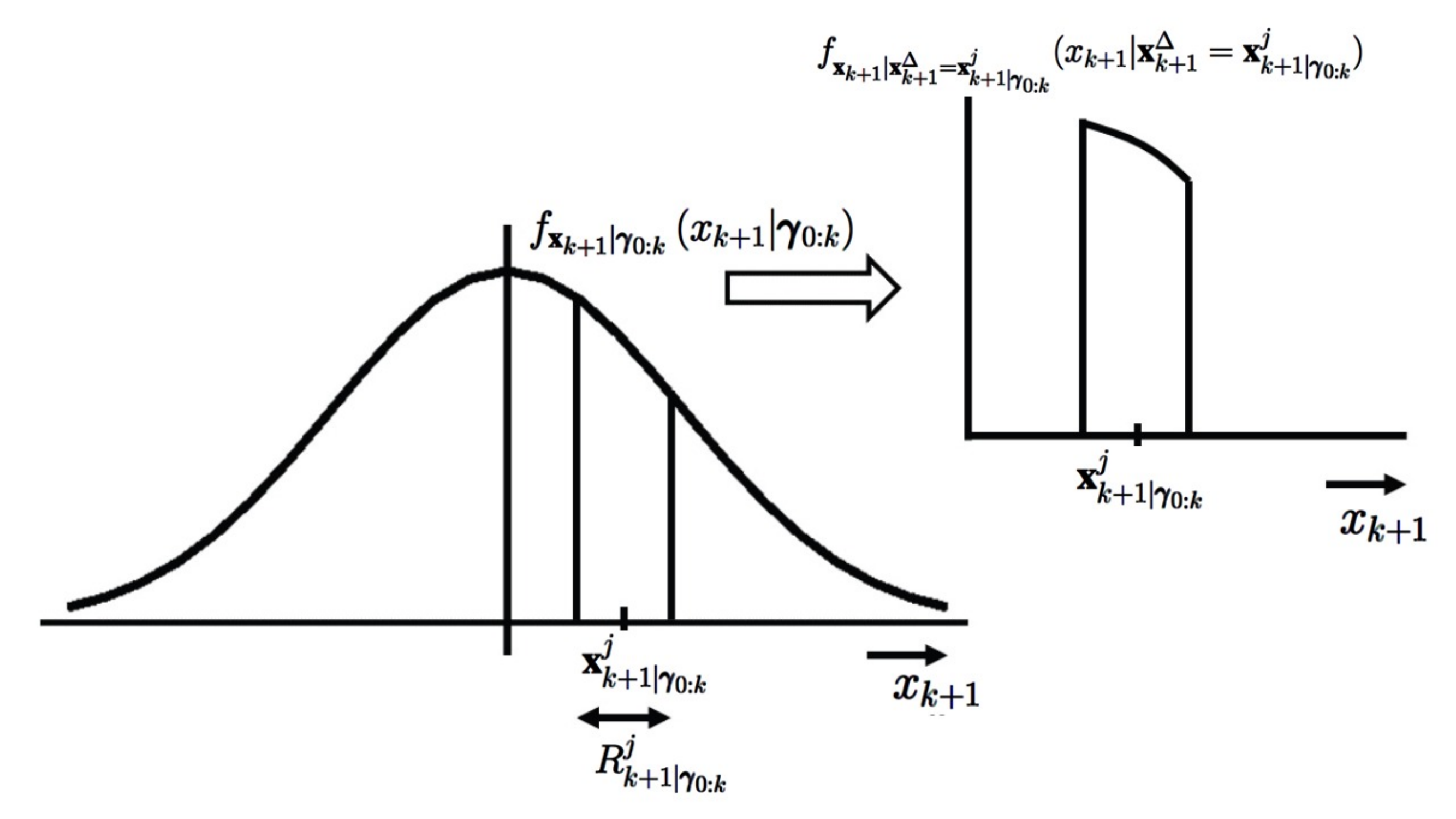}
\caption{The conditional density function of the state conditioned on a  quantization region, and the sequence of intermittence random variables}
\label{iLQG}
\end{figure}
That completes the proof. \hspace{12.3cm}$\blacksquare$

\subsection{Proof of Lemma \ref{lemma3}}
\label{proof3}
 Let the initial step size for quantization be $\Delta_0$. From Shannon's capacity theorem \cite{shannon1948mathematicalf,cover1999elements}, $\frac{2^{h({\bf x}_0)}}{\Delta_0} \le 2^{\mathcal{C}}$
    $\Rightarrow \Delta_0 = \frac{\sqrt{2\pi e}\sigma_0}{2^{\mathcal{C}}} + {\epsilon}$, where ${\epsilon}$ is a small positive constant. Similarly, for $(k+1)^{th}$ time step, $\Delta_{k+1} = \frac{\sqrt{2\pi e}\sigma_{k+1}}{2^{\mathcal{C}}} + {\epsilon}$.
    Conditioning the quantization step-sizes with respect to the intermittence random sequences , the one step conditional update equation becomes
    \begin{eqnarray}
    \label{cond_quant_bern_scal_eqn}
      \nonumber  &&{\pmb \Delta}_{k+1|{\pmb \gamma}_{0:k}}^2 = \frac{2\pi e {\pmb \sigma}_{k+1|{\pmb \gamma}_{0:k}}^2}{2^{2\mathcal{C}}} + \epsilon 
        =\frac{2\pi e}{2^{2\mathcal{C}}}\left[{\pmb \xi}_{k}^2{\pmb \sigma}_{k|{\pmb \gamma}_{0:k-1}}^2 + b^2 l^2\left(1 - {\pmb \gamma}_k\right)\frac{{\pmb\Delta}_{k|{\pmb \gamma}_{0:k-1}}^2}{12} + \sigma_{\bf w}^2\right] + \epsilon \\
     %   &=& {\pmb \xi}_{k}^2\left(\Delta_{k}^2 - \epsilon \right) + \eta L\left(1 - \gamma_k\right) \Delta_{k}^2 + \eta \sigma_{w}^2 + \epsilon \\
        &=& {\pmb \Delta}_{k|{\pmb \gamma}_{0:k-1}}^2 \left[{\pmb \xi}_{k}^2 + \eta G\left(1 - {\pmb\gamma}_k\right)\right] + \eta \sigma_{\bf w}^2 + \epsilon \left(1 - {\pmb\xi}_{k}^2\right) 
    \end{eqnarray}
    
    where, $\eta = \frac{2\pi e} {2^{2\mathcal{C}}}$, $G = \frac{b^2 l^2}{12}$, and $\epsilon$ is a small positive constant.
    Taking expectation over ${\pmb \gamma}_{0:k}$, the expected one step quantization update equation is obtained. 
    \begin{eqnarray*}
        \Delta_{k+1}^2 &=& \Delta_{k}^2\left[\omega^2 + \eta G(1 - p)\right] + \eta \sigma_{w}^2 + \epsilon\left(1 - \omega^2\right)
    \end{eqnarray*}
    That concludes the proof. \hspace{12.5cm}$\blacksquare$

\subsection{Proof of Theorem \ref{theorem4}}
\label{proof4}
\begin{enumerate}
    \item  For $\mathcal{L}_2$ stabilization of the system of section \ref{scalar_bernoulli}, and assuming $\mathcal{C}\rightarrow\infty$,  Eqn. \ref{lemma1_eqn} from Lemma \ref{lemma1} states that the asymptotic state variance $\lim_{n\rightarrow\infty}\sigma_{n}^2 < \infty$ if and only if $\omega^2 < 1$. So, the necessary and sufficient condition is:
        \begin{eqnarray*}
            \alpha^2 p + (\alpha-bl)^2 (1 - p) < 1 \\
            \Rightarrow p < \frac{1 - (\alpha-bl)^2}{\alpha^2 - (\alpha-bl)^2}
        \end{eqnarray*}
    Moreover, the asymptotic state variance will be:
    \begin{eqnarray*}
        \lim_{n\rightarrow\infty}\sigma_{n}^2 =\lim_{n\rightarrow\infty}\left[ \omega^{2(n+1)}\sigma_{0}^2 + \left(\frac{1 - \omega^{2(n+1)}}{1 - \omega^2}\right)\sigma_{\bf w}^2\right]
        = \frac{\sigma_{\bf w}^2}{(1 - \omega^2)}
    \end{eqnarray*}
 
 \item Now consider $\frac{1}{2}\log\left(\frac{\pi e \alpha^2}{6}\right)<\mathcal{C}<\infty$. Conditioning Eqn. \ref{cond_var_bern_scalar_eqn} on the intermittence random vector ${\pmb\gamma}_{0:k-1}$, the following random equation is obtained. 
\begin{eqnarray*}
\label{cond_var_bern_scal_eqn}
    {\pmb \sigma}_{k+1|{\pmb\gamma}_{0:k}}^2 &=& {\pmb \xi}_{k}^2{\pmb\sigma}_{k|{\pmb\gamma}_{0:k-1}}^2 + G(1 - {\pmb\gamma}_k) {\pmb\Delta}_{k|{\pmb\gamma}_{0:k-1}}^2+ \sigma_{\bf w}^2 
\end{eqnarray*}

Moreover, using Eqn. \ref{cond_quant_bern_scal_eqn} for the $k^{th}$ time step, the following equation is obtained. 
\begin{eqnarray*}
    {\pmb \Delta}_{k|{\pmb\gamma}_{0:k-1}}^2 = \eta {\pmb\sigma}_{k|{\pmb\gamma}_{0:k-1}}^2 + \epsilon
\end{eqnarray*}
    Combining the above two equations, the following recursion is obtained. 
        \begin{eqnarray*}
            {\pmb \sigma}_{k+1|{\pmb \gamma}_{0:k}}^2 &=& \left[{\pmb\xi}_{k}^2 + \eta G\left(1 - {\pmb\gamma}_k\right)\right]{\pmb\sigma}_{k|{\pmb \gamma}_{0:k-1}}^2 
            + \left[\sigma_{\bf w}^2 + G\left(1 - \gamma_k\right)\epsilon\right]
        \end{eqnarray*}
Taking expectation with respect to ${\pmb \gamma}_{0:k}$, and extending the recursion up to the initial condition, the variance formula is obtained. 
 %       \begin{eqnarray*}
  %          \sigma_{k+1}^2 &=& \left[\omega^2 + \eta L(1 - p)\right]\sigma_{k}^2 + \left[\sigma_{w}^2 + L(1 - p)\epsilon\right] 
   %     \end{eqnarray*}
    %    And,
        \begin{eqnarray*}
            \sigma_{n}^2 = \left[\omega^2 + \eta G(1 - p)\right]^n\sigma_{0}^2  
            + \sum_{k=0}^{n-1}\left[\sigma_{\bf w}^2 + G(1-p)\epsilon \right] \left[\omega^2 + \eta G(1-p)\right]^k
        \end{eqnarray*}
        \label{var_Q}
    Clearly the necessary and sufficient condition for  $\lim_{n\rightarrow\infty} \sigma_n^2 <\infty$ is given by: 
    \begin{eqnarray}
    \label{scal_bern_p_cond_Eq}
        \left[\omega^2 + \eta G(1-p)\right]< 1
    \end{eqnarray}

Here, it will be imperative to observe the influence of the finite capacity $\mathcal{C}$ on the closed loop even in absence of any intermittent unstable open loop. Also consider an infinite control cost which will make the closed loop gain very small. With these ideal conditions of $p=0$, and  $(\alpha-bl)\rightarrow 0$, Eqn. \ref{scal_bern_p_cond_Eq} will become:
\begin{eqnarray}
\label{scal_bern_C_cond_eq}
   \nonumber \eta G <1 \Rightarrow \frac{2 \pi e}{2^{2\mathcal{C}}}\frac{b^2l^2}{12} <1
    \Rightarrow \frac{\pi e \alpha^2}{6}<2^{2\mathcal{C}}\\
    \Rightarrow \mathcal{C}> \frac{1}{2}\log{\left(\frac{\pi e \alpha^2}{6}\right)}
\end{eqnarray}

Clearly, in non-ideal condition, the necessary (not sufficient) condition for $\mathcal{L}_2$ stabilization of the system is given in Eqn. \ref{scal_bern_C_cond_eq}. For large enough $\mathcal{C}$, the upper bound of $p$ for $\mathcal{L}_2$ stabilization of the system is thus obtained by rearranging Eqn. \ref{scal_bern_p_cond_Eq}.
   \begin{eqnarray}
         p < \frac{1 - (\alpha-bl)^2 - \eta G}{\alpha^2 - (\alpha-bl)^2 -  \eta G}    
        \end{eqnarray}

        Moreover, the asymptotic state variance is given by:
        \begin{eqnarray*}
            \lim_{n \rightarrow \infty}\sigma_{n}^2 &=& \frac{\sigma_{\bf w}^2 + G(1 - p)\epsilon}{1 - \left(\omega^2 + \eta G(1 - p)\right)} 
        \end{eqnarray*}    
\item If $\mathcal{C} \le \frac{1}{2}\log{\left(\frac{\pi e \alpha^2}{6}\right)}$, from Eqn. \ref{scal_bern_p_cond_Eq}, the necessary condition breaks down, and even in the ideal condition of $p=0$, and $\alpha = bl$, the system goes $\mathcal{L}_2$ unstable. 
That concludes the proof.   \hspace{4.1cm}$\blacksquare$
\end{enumerate}

\subsection{Proof of Lemma \ref{lemma5}}
\label{proof5}
The stochastic system of section \ref{vect_bern_sys} with control law ${\bf u}_k=-L{\bf x}_k$ is given by:
\begin{eqnarray*}
        {\bf x}_{k+1} &=& {\pmb \gamma}_k A {\bf x}_k + \left(1 - {\pmb \gamma}_k\right)\left(A-BL\right) {\bf x}_k + {\bf w}_k \\
        &=& {\pmb \Lambda}_k {\bf x}_k + {\bf w}_k 
\end{eqnarray*}
Where, ${\pmb \Lambda}_k = {\pmb \gamma}_k A+ \left(1 - {\pmb \gamma}_k\right)\left(A-BL\right) $ is a random matrix with every element a Bernoulli random variable. The one step conditional state covariance matrix equation is given by:
\begin{eqnarray*}
{\bf P}_{k+1|{\pmb \gamma}_k} = \mathbb{E}\left({\bf x}_{k+1}{\bf x}'_{k+1}|{\pmb \gamma}_k\right) = {\pmb\Lambda}_k P_k{\pmb\Lambda}'_k +W
\end{eqnarray*}
where, $P_k=\mathbb{E}\left({\bf x}_{k}{\bf x}'_{k}\right)$ is the state covariance matrix at $k^{th}$ time step. 
Conditioning on ${\pmb \gamma}_{0:k-1}$, the equation becomes:
\begin{eqnarray}
\label{vec_bern_cov_1_eq}
{\bf P}_{k+1|{\pmb \gamma}_{0:k}} = {\pmb \Lambda}_k {\bf P}_{k|{\pmb \gamma}_{0:k-1}}{\pmb\Lambda}'_k +W
\end{eqnarray}
Expanding the recursion till the initial condition, the following expression is obtained:
\begin{eqnarray}
    {\bf P}_{k+1|{\pmb \gamma}_{0:k}}=\left(\prod_{j=k}^{0}{\pmb\Lambda}_j\right) {P}_0\left(\prod_{j=0}^{k}{\pmb\Lambda}'_j\right)
    +\left[\sum_{j=0}^{k-1}\left\{\left(\prod_{i=k}^{j+1}{\pmb\Lambda}_i\right)W\left(\prod_{i=j+1}^{k}{\pmb\Lambda}'_i\right)\right\}+W\right]
\end{eqnarray}.
That concludes the proof. \hspace{12cm}  $\blacksquare$

\subsection{Proof of Lemma \ref{lemma6}}
\label{proof6}
The Bernoulli random sequence  $\{{\pmb\gamma}_0, {\pmb\gamma}_1, \dots {\pmb\gamma}_n,\dots \}$ is stationary with $\mathbb{E}({\pmb\gamma}_k) = p$.  From Eqn \ref{vec_bern_cov_1_eq}, it can be seen that
\begin{eqnarray*}
{\bf P}_{k+1|{\pmb \gamma}_k} = {\pmb\Lambda}_k P_k {\pmb\Lambda}'_k +W
= {\pmb\gamma}_k AP_{k} A'+(1-{\pmb\gamma}_k) (A-BL)P_{k}(A-BL)'+W
\end{eqnarray*}
Where, $P_{k}=\mathbb{E}\left({\bf x}_{k}{\bf x}_{k}'\right)$.
Taking expectation with respect to ${\pmb\gamma}_k$, the desired form (Eqn.\ref{vec_bern_cov_eq}) is obtained. That concludes the proof. \hspace{14.4cm} $\blacksquare$
\subsection{Proof of Theorem \ref{theorem7}}
\label{proof7}
The dynamics of the state covariance matrix $P_k$ as discussed in lemma \ref{vector_bernoulli_covariance_expected} (Eqn. \ref{vec_bern_cov_eq}) can be visualized as a flow in the open convex cone of positive definite matrices $(M_n^{+})$. Obviously the set is not a metric space. However, the open set can be completed by including the singular positive semi-definite matrices to form the closed cone $(M_n)$. Since the set is not compact, not all flows lead to a fixed point. However, if a contraction operator can be constructed to define the dynamics, it will asymptotically converge to a fixed point. For any $p\in [0,1]$, define the operator $\mathcal{T}_p:M_n\rightarrow M_n$ as:
\begin{eqnarray*}
\mathcal{T}_p(P) = pAPA'+(1-p)(A-BL)P(A-BL)'
\end{eqnarray*}
Then the state covariance update equation can be written as:
\begin{eqnarray*}
P_{k+1} = \mathcal{T}_p(P_{k})+W
\end{eqnarray*}

\begin{figure}[h]
\centering
\includegraphics[width=1.5in]{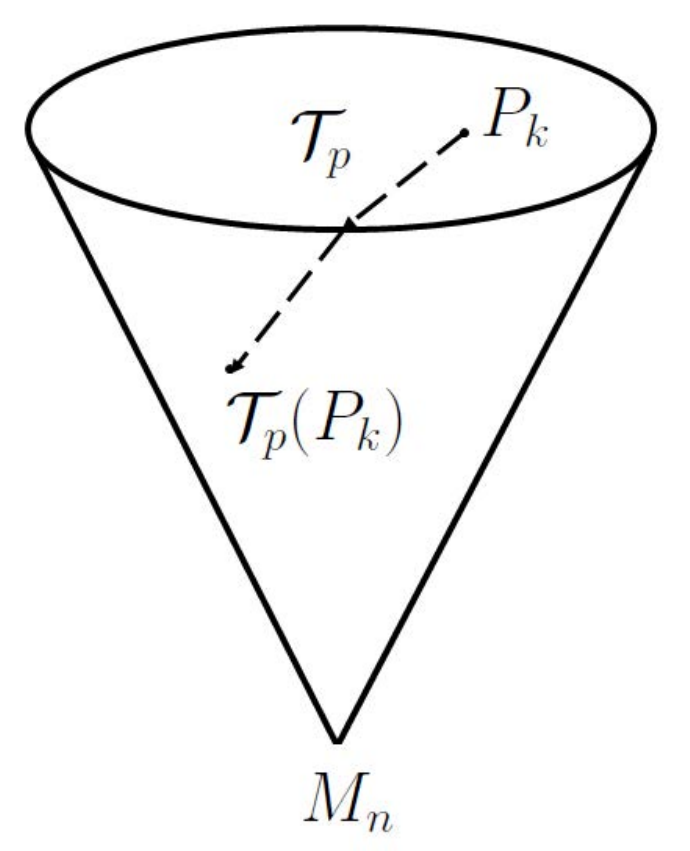}
\caption{The contraction operator $\mathcal{T}_p$ in the cone of positive semi-definite matrices $M_n$}
\label{cone}
\end{figure}

Since $W\in M_n^+$, and $\mathcal{T}_p(P_{k}) \in M_n$, it implies that  $P_{k+1} \in M_n^+$. Clearly, $\lim_{k\rightarrow \infty} P_{k}$ will be unique and bounded if and only if $\mathcal{T}_p$ is a contraction or equivalently $\norm{\mathcal{T}_p(P)}_2<\norm{P}_2$ for all $P\in M_n$. Now,
\begin{eqnarray}
\nonumber\norm{\mathcal{T}_p(P)}_2 &=& \norm{pAPA'+(1-p)(A-BL)P(A-BL)'}_2\\
&\le& \norm{pAPA'}_2+\norm{(1-p)(A-BL)P(A-BL)'}_2\\
&=& p\norm{APA'}_2+(1-p)\norm{(A-BL)P(A-BL)'}_2\\
&\le& p\norm{A}_2^2\norm{P}_2+(1-p)\norm{A-BL}_2^2\norm{P}_2\\
\nonumber&=& \norm{P}_2 \left(p\norm{A}_2^2+(1-p)\norm{A-BL}_2^2\right)
\end{eqnarray}
Here Eqn.41 is obtained from the triangle inequality of spectral norm, Eqn. 42 is obtained from the scalar multiplication of spectral norm, and Eqn. 43 is obtained from the submultiplicative inequality of spectral norm. Therefore, the necessary and sufficient condition for ${\mathcal{T}_p}$ to be a contraction is given by:
$\left[p\norm{A}_2^2+(1-p)\norm{A-BL}_2^2\right]<1$. By rearrangement, the expression of the theorem (Eqn. \ref{vec_markov_iff_l2}) is obtained. That concludes the proof. \hspace{3cm}$\blacksquare$

\subsection{Proof of Lemma \ref{lemma8}}
\label{proof8}
When ${\pmb \gamma}_k = 1$, the open loop system is given by: ${\bf x}_{k+1} = A {\bf x}_k+{\bf w}_k$, and for ${\pmb \gamma}_k = 0$, the loop is closed using the control signal ${\bf u}_k = -L {\bf x}_k^\Delta$. Here, ${\bf x}_k^\Delta$ is the quantized state at time $k$. The quantization error is ${\bf e}_k = {\bf x}_k-{\bf x}_k^\Delta$. So, the closed loop dynamics is given by:
    \begin{eqnarray*}
        {\bf x}_{k+1} &=& A {\bf x}_k+B(-L{\bf x}_k^\Delta)+{\bf w}_k\\
        &=& (A-BL) {\bf x}_k+BL{\bf e}_k+{\bf w}_k
    \end{eqnarray*}
    Hence the stochastic dynamics based on the Bernoulli switching random variable ${\pmb\gamma}_k$ is given by:
    \begin{eqnarray*}
        {\bf x}_{k+1} = {\pmb\Lambda}_k{\bf x}_k+(1-{\pmb\gamma}_k)BL{\bf e}_k+{\bf w}_k
    \end{eqnarray*}
    Where, ${\pmb\Lambda}_k = \left\{{\pmb\gamma}_kA+(1-{\pmb\gamma}_k)(A-BL)\right\}$. The one step update of the conditional state covariance matrix at time $(k+1)$ is given by:
    \begin{eqnarray*}
        &&{\bf P}_{k+1|{\pmb\gamma}_{0:k}} = \\&&{\pmb\gamma}_k A {\bf P}_{k|{\pmb \gamma}_{0:k-1}}A'
        + (1-{\pmb\gamma}_k)(A-BL){\bf P}_{k|{\pmb \gamma}_{0:k-1}}(A-BL)'
        + (1-{\pmb\gamma}_k)BL\mathbb{E}({\bf e}_k{\bf e}_k'|{\pmb\gamma}_{0:k-1})L'B'+W
    \end{eqnarray*}
    Since $m<n$, the matrix $BL$ is singular, and hence the innovation contribution (differential entropy) of the quantization error vector ${\bf e}_k$ to the state differential entropy is $-\infty$, resulting in the volume of the corresponding typical support set being zero. Hence, this method cannot be used to relate the state covariance matrix and the quantization step-size like Lemma \ref{lemma2}. That concludes the proof. \hspace{8.7cm} $\blacksquare$

\subsection{Proof of Theorem \ref{theorem9}}
\label{proof9}
For Gaussian random vector ${\bf x}_k$ with conditional covariance matrix ${\bf P}_{k|{\pmb\gamma}_{0:k-1}}$, the conditional differential entropy is given by:
\begin{eqnarray}
\label{vec_bern_cond_de_eq}
   \nonumber {\bf h}\left({\bf x}_k|{\pmb \gamma}_{0:k-1}\right) &=& \frac{1}{2}\log\left(\left(2\pi e\right)^{n} \left|{\bf P}_{k|{\pmb\gamma}_{0:k-1}}\right|\right)\\
\Rightarrow \left|{\bf P}_{k|{\pmb\gamma}_{0:k-1}}\right|^\frac{1}{n} &=& \left(\frac{1}{2 \pi e}\right) 2^{\left[\frac{2}{n} {\bf h}\left({\bf x}_k|{\pmb \gamma}_{0:k-1}\right)\right]}
\end{eqnarray}
From lemma \ref{lemma5}, the one step update rule for the conditional covariance is obtained. 
\begin{eqnarray}
\label{vec_bern_cov_cond}
    {\bf P}_{k+1|{\pmb\gamma}_{0:k}} ={\pmb\gamma}_k A {\bf P}_{k|{\pmb\gamma}_{0:k-1}}A'
    +(1-{\pmb\gamma}_k)(A-BL) {\bf P}_{k|{\pmb\gamma}_{0:k-1}}(A-BL)'+W
\end{eqnarray}
Applying Minkowskii's ineqaulity on Eqn. \ref{vec_bern_cov_cond}, the following inequality is obtained. 
\begin{eqnarray}
\left|{\bf P}_{k+1|{\pmb\gamma}_{0:k}}\right|^\frac{1}{n} \ge 
\left[{\pmb\gamma}_k |A|^\frac{2}{n}+(1-{\pmb\gamma}_k) |A-BL|^\frac{2}{n} \right]\left|{\bf P}_{k|{\pmb\gamma}_{0:k-1}}\right|^\frac{1}{n}+|W|^\frac{1}{n}
\end{eqnarray}
Using conditional differential entropy from Eqn. \ref{vec_bern_cond_de_eq} in this form gives a version of the Entropy Power Inequality (EPI) \cite{shannon1948mathematicalf}:
\begin{eqnarray}
\label{vec_bern_epi}
 2^{\left[\frac{2}{n} {\bf h}\left({{\bf x}_{k+1}}|{\pmb \gamma}_{0:k}\right)\right]} \ge 2^{\left[\frac{2}{n} h\left({\bf w}_k\right)\right] } + 2^{\left[\frac{2}{n}{\bf h}\left({\bf x}_{k}|{\pmb \gamma}_{0:k-1}\right)\right]} \left[{\pmb\gamma}_k |A|^{\frac{2}{n}}+(1-{\pmb\gamma}_k) |A-BL|^\frac{2}{n} \right]
\end{eqnarray}
The quantization step-size depends on the typical conditional support set of the state vector at every instance. If the conditional quantization step-size per dimension at time instance $k+1$ is ${\pmb \Delta}_{k+1|{\pmb \gamma}_{0:k}}$, and the typical conditional support set cardinality is $2^{\left[\frac{2}{n}{\bf h}\left({\bf x}_{k+1}|{\pmb \gamma}_{0:k}\right)\right]}$, it is deduced that:
\begin{eqnarray}
\label{vec_bern_quant_final}
    \log\left[\frac{2^{\left[\frac{2}{n}{\bf h}\left({\bf x}_{k+1}|{\pmb \gamma}_{0:k}\right)\right]}}{{\pmb \Delta}^n_{k+1|{\pmb \gamma}_{0:k}}}\right]<\mathcal{C}
\end{eqnarray}
Combining the above two inequalities (\ref{vec_bern_epi} and \ref{vec_bern_quant_final}), the theorem statement is obtained completing the proof. \hspace{0.7cm} $\blacksquare$

\subsection{Proof of Lemma \ref{lemma10}}
\label{proof10}
For Markov intermittence, the stochastic system can be described by, ${\bf x}_{k+1} = {\pmb\xi}_k {\bf x}_k + {\bf w}_k$, where, the initial state ${\bf x}_0\sim \mathcal{N}(0,\sigma_0^2)$, and ${\pmb \xi}_k = \alpha {\pmb \gamma}_k + (\alpha-bl)\left(1 - {\pmb\gamma}_k\right)$. Since ${\pmb\gamma}_k$ is a Markov chain with transition probability matrix (TPM) $T$, ${\pmb \xi}_k$ is also a binary Markov chain with the same TPM $T$ with the two states being the open loop gain $\alpha$, and the closed loop gain $(\alpha-bl)$. Clearly, ${\pmb \xi}_k^2 = \alpha^2 {\pmb\gamma}+(\alpha-bl)^2(1-{\pmb\gamma}_k)$, and $\omega_k^2 = \mathbb{E}({\pmb\xi}_k^2)=\alpha^2\pi_k+(\alpha-bl)^2(1-\pi_k)$. The probability state vector $\zeta_k = [(1-\pi_k), \pi_k]$ of the Markov chain follows the dynamics $\zeta_{k+1}=\zeta_k T$ with the initial condition $\zeta_0$. 
The conditional state variance ${\pmb \sigma}_{k+1|{\pmb\gamma}_{0:k}}^2$ at the $(k+1)^{th}$ time step depends only on the initial condition, and the sequence ${\pmb\gamma}_{0:k}$, the formulation is identical to the derivation under the Bernoulli intermittence assumption as shown in section \ref{proof1}. Therefore, the repetition is omitted to prove Eqn. \ref{scal_markov_cond_var_eq}. However, the sequence ${\pmb\gamma}_{0:k}$ for a Markov switching scheme is not a sequence of independent and/or identically distributed random variables like the Bernoulli example. 
Hence, taking expectation of Eqn. \ref{scal_markov_cond_var_eq} with respect to ${\pmb\gamma}_{0:k}$ will not reduce the formula like the Bernoulli one. Instead the form of Eqn. \ref{scal_markov_var_eq} is obtained. That concludes the proof. \hspace{7.5cm}
$\blacksquare$

\subsection{Proof of Lemma \ref{lemma11}}
\label{proof11}
The first part of the proof is similar to that of the Bernoulli intermittence model as shown in section \ref{proof2}. The one step conditional variance update equation (Eqn. \ref{cond_var_markov_scalar_eqn}) is identical to Eqn. \ref{cond_var_bern_scalar_eqn}. Hence, the proof is omitted. Now, taking expectation of Eqn. \ref{cond_var_markov_scalar_eqn} with respect to ${\pmb\gamma}_{0:k}$ results in Eqn. \ref{var_markov_scal_eqn}.
That concludes the proof. \hspace{3.3cm}$\blacksquare$

\subsection{Proof of Lemma \ref{lemma12}}
\label{proof12}
The proof of Eqn. \ref{scal_markov_cond_quant_eq} is exactly same as Eqn. \ref{scal_bern_cond_quant_eq} which is elaborated in section \ref{proof3}. Next, taking expectation of Eqn. \ref{scal_markov_cond_quant_eq} with respect to ${\pmb\gamma}_{0:k}$ results in Eqn. \ref{scal_markov_quant_eq}. That concludes the proof. \hspace{6.1cm}$\blacksquare$ 

\subsection{Proof of Theorem \ref{theorem13}}
\label{proof13}
\begin{enumerate}
        \item  The necessary and sufficient condition for $\mathcal{L}_2$ stabilization (i.e., $\lim_{k\rightarrow\infty}\sigma_k^2<\infty$) of the system under the assumption of $\mathcal{C}\rightarrow\infty$ can be obtained from Eqn. \ref{scal_markov_var_eq} as:
        $\lim_{k\rightarrow\infty}\prod_{j=0}^{k}\omega_{j}^2 < \infty$. This will be satisfied if and only if $\lim_{k\rightarrow\infty}\omega_{k}^2 < 1$. Now, $\omega_{k}^2 = \alpha^2 \pi_{k} + (\alpha-bl)^2\left(1 - \pi_{k}\right)$, and $\zeta_{\infty}=[(1-\pi_\infty),\pi_\infty]=\lim_{k\rightarrow\infty}\zeta_{k}$ is the steady state distribution of the intermittent switch Markov Chain. The steady state distribution $\zeta_\infty$ is the left eigenvector of the transition probability matrix $T$ with unity eigenvalue. So, $\pi_\infty$ is obtained by solving the eigen-equation $\zeta_\infty = \zeta_\infty T$, and is given by: $\pi_\infty = \frac{p}{p+q}$. 
        Therefore, the condition is:
        \begin{eqnarray*}
        \lim_{k\rightarrow\infty}\omega_k^2 = \alpha^2\pi_\infty+ (\alpha-bl)^2(1-\pi_\infty) < 1  
        \end{eqnarray*}
        That provides the upper bound of the ratio of the two transition probability parameters of the Markov chain for $\mathcal{L}_2$ stabilization of the system under the assumption $\mathcal{C\rightarrow\infty}$:
        \begin{eqnarray*}
            \frac{p}{q} < \frac{1 - (\alpha-bl)^2}{\alpha^2 - 1} 
        \end{eqnarray*}
        Now, if $\lim_{k\rightarrow\infty}\omega_{k}^2 < 1$, then $\lim_{k\rightarrow\infty}\prod_{j=0}^{k}\omega_{j}^2 \rightarrow 0$.
        Therefore the asymptotic behavior of the state variance can be obtained from Eqn. \ref{scal_markov_var_eq} by taking limit $k\rightarrow\infty$.
        
        \begin{eqnarray*}
          \lim_{k\rightarrow\infty} \sigma_k^2   &=&  \lim_{k\rightarrow\infty}\left(\frac{\sigma_{\bf w}^2}{1 - \omega_k^2}\right) \\
          &=& \frac{\sigma_{\bf w}^2}{1 - \left[\alpha^2\pi_{\infty} + (\alpha-bl)^2\left(1 - \pi_{\infty}\right)\right]} \\
            &=& \left[\frac{\left(1 + \frac{p}{q}\right)}{1 - (\alpha-bl)^2 - \left(\alpha^2 - 1\right)\frac{p}{q}}\right]\sigma_{\bf w}^2
        \end{eqnarray*}
        \item For finite capacity system ($\mathcal{C}<\infty$), Eqn. \ref{cond_var_markov_scalar_eqn} gives the one step conditional state variance update equation under the quantized scheme. 
        Also, from Shannon's capacity theorem \cite{shannon1948mathematicalf,cover1999elements}, the relationship of state variance with the quantization step-size is obtained as:
        \begin{eqnarray*}
            {\pmb \Delta}_{k|{\pmb\gamma}_{0:k-1}}^2 = \eta {\pmb \sigma}_{k|{\pmb\gamma}_{0:k-1}}^2+\epsilon
        \end{eqnarray*}
    Combining this result with Eqn. \ref{cond_var_markov_scalar_eqn}, the following recursion is obtained:
    \begin{eqnarray*}
        &&{\pmb\sigma}_{k+1|{\pmb\gamma}_{0:k}}^2 = \\
       &&\left[{\pmb\xi}_k^2+\eta G (1-{\pmb\gamma}_k)\right]{\pmb\sigma}_{k|{\pmb\gamma}_{0:k-1}}^2+\left[\sigma_{\bf w}^2+G(1-{\pmb\gamma}_k)\epsilon\right]
    \end{eqnarray*}
    Taking expectation with respect to ${\pmb\gamma}_{0:k}$ results in 
    \begin{eqnarray*}
        \sigma_{k+1}^2 = [\omega_k^2+\eta G (1-\pi_k)]\sigma_k^2+[\sigma_{\bf w}^2+G(1-\pi_k)\epsilon]
    \end{eqnarray*}
    The necessary and sufficient condition for $\lim_{k\rightarrow\infty} \sigma_k^2<\infty$ is $\lim_{k\rightarrow\infty} [\omega_k^2+\eta G (1-\pi_k)]<1$.
    The steady state distribution of the intermittent switch Markov chain is $\zeta_\infty = [(1-\pi_\infty), \pi_\infty] = [\frac{q}{p+q}, \frac{p}{p+q}]$. Hence, the necessary and sufficient condition for $\mathcal{L}_2$ stabilizability of the system is given by:
    \begin{eqnarray*}
        \alpha^2\pi_\infty+[(\alpha-bl)^2 + \eta G](1-\pi_\infty)<1
    \end{eqnarray*}
    The condition in terms of the ratio of the transition probability parameters is given by:
        \begin{eqnarray*}
        \frac{p}{q} < \frac{1 - (\alpha-bl)^2 - \eta G}{\alpha^2 - 1}
        \end{eqnarray*}

Then the asymptotic state variance is given by:
        
         \begin{eqnarray*}
          \lim_{k\rightarrow\infty} \sigma_k^2 &=& \frac{\sigma_{\bf w}^2 + G\left(1 - \pi_{\infty}\right)\epsilon}{1 - \left(\omega_{\infty}^2 + \eta G\left(1 - \pi_{\infty}\right)\right)} \\
          &=& \frac{\left(1 + \frac{p}{q}\right)\sigma_{\bf w}^2 + \epsilon G}{\left(1 + \frac{p}{q}\right) - \left[\alpha^2 \frac{p}{q} + (\alpha-bl)^2 + \eta G\right]}
        \end{eqnarray*}
    This proves Eqn. \ref{scal_markov_iff_fin}.  Moreover, in idealized condition of $p=0$ (i.e., the transition probability of the switch from `ON' to `OFF' state is zero), $q=1$ (i.e., the transition probability of the switch from `OFF' to `ON' state is unity), and $(\alpha-bl)\rightarrow 0$ (i.e., the closed loop gain is extremely small, which is feasible with infinite control cost), the condition for $\mathcal{L}_2$ stability of the system becomes: $\eta G <1$. That gives the lower bound of the capacity for $\mathcal{L}_2$ stabilization. Assuming $\alpha = b l$, 
    \begin{eqnarray*}
        \frac{2\pi e}{2^{2\mathcal{C}}}\frac{b^2l^2}{12}<1\\
        \Rightarrow \mathcal{C}>\frac{1}{2}\log\left(\frac{\alpha^2}{6\pi e}\right)
    \end{eqnarray*}
    \item If $\mathcal{C}\le\frac{1}{2}\log\left(\frac{\alpha^2}{6\pi e}\right)$, even in the idealized situation the state variance will diverge. The system will be always $\mathcal{L}_2$ unstable. 
    \end{enumerate}
    Thus the proof is concluded. \hspace{12cm}$\blacksquare$

\subsection{Proof of Lemma \ref{lemma14}}
\label{proof14}
The system as discussed in section \ref{vector_markov} is given by:
\begin{eqnarray*}
        {\bf x}_{k+1} &=& {\pmb\gamma}_k A {\bf x}_k + \left(1 - {\pmb\gamma}_k\right)\left(A-BL\right) {\bf x}_k + {\bf w}_k \\
        &=& {\pmb\Lambda}_k {\bf x}_k + {\bf w}_k 
\end{eqnarray*}
Hence, ${\bf P}_{k+1|{\pmb\gamma}_{0:k}}= {\pmb\Lambda}_k {\bf P}_{k|{\pmb\gamma}_{0:k-1}}{\pmb\Lambda}'_k + W$. Eqn. \ref{vec_markov_cond_cov_eq} is obtained by expanding the recursion till the initial condition . That concludes the proof. \hspace{12.3cm}$\blacksquare$

\subsection{Proof of Lemma \ref{lemma15}}
\label{proof15}
The Markov sequence  $\{{\pmb\gamma}_0, {\pmb\gamma}_1, \dots {\pmb\gamma}_n,\dots \}$ is ergodic with $P({\pmb\gamma}_k = 1) = \pi_k$.  From Lemma \ref{lemma14}, 
\begin{eqnarray*}
{\bf P}_{k+1|{\pmb\gamma}_{0:k}} =
 {\pmb\gamma}_k A{\bf P}_{k|{\pmb\gamma}_{0:k-1}} A'
 +(1-{\pmb\gamma}_k) (A-BL){\bf P}_{k|{\pmb\gamma}_{0:k-1}}(A-BL)'+W
\end{eqnarray*}
Eqn. \ref{vec_markov_cov_eq} is obtained by taking expectation of the above equation with respect to ${\pmb\gamma}_{0:k}$. That concludes the proof. \hspace{.3cm}$\blacksquare$

\subsection{Proof of Theorem \ref{theorem16}}
\label{proof16}
The dynamics of the state covariance matrix $P_k$ as given in Eqn. \ref{vec_markov_cov_eq} in Lemma \ref{lemma15} can be visualized as a flow in the open convex cone of positive definite matrices $(M_n^{+})$. Obviously the set is not a metric space. However, the open set can be completed by including the singular positive semi-definite matrices to form the closed cone $(M_n)$. Since the set is not compact, not all flows lead to a fixed point. Therefore, if a sequence of contraction operators can be constructed to define the dynamics of the covariance matrix, it will asymptotically converge to a fixed point in $M_n^{+}$. For the sequence $\{\pi_k\in [0,1]\} , \forall k\in \mathbb{N}$, define a sequence of operators $\mathcal{T}{_k}:M_n\rightarrow M_n$ as:
\begin{eqnarray*}
\mathcal{T}_k(P) = \pi_k APA'+(1-\pi_k)(A-BL)P(A-BL)'
\end{eqnarray*}
Then the state covariance update equation can be written as:
\begin{eqnarray*}
P_{k+1} = \mathcal{T}_k(P_{k})+W
\end{eqnarray*}
Since $W\in M_n^+$, and $\mathcal{T}_k(P_{k}) \in M_n$, it implies that  $P_{k+1} \in M_n^+$. Clearly $\lim_{k\rightarrow \infty} P_{k}$ will be unique and bounded if and only if $\lim_{k\rightarrow\infty}\mathcal{T}_k$ is a contraction or equivalently $\lim_{k\rightarrow\infty}\norm{\mathcal{T}_k(P)}_2<\norm{P}_2$ for all $P\in M_n$. Now,
\begin{eqnarray}
\nonumber\lim_{k\rightarrow\infty}\norm{\mathcal{T}_k(P)}_2 &=& \norm{\pi_\infty APA'+(1-\pi_\infty)(A-BL)P(A-BL)'}_2\\
&\le& \norm{\pi_\infty APA'}_2+\norm{(1-\pi_\infty)(A-BL)P(A-BL)'}_2\\
&=& \pi_\infty\norm{APA'}_2+(1-\pi_\infty)\norm{(A-BL)P(A-BL)'}_2\\
&\le& \pi_\infty\norm{A}_2^2\norm{P}_2+(1-\pi_\infty)\norm{A-BL}_2^2\norm{P}_2\\
\nonumber&=& \norm{P}_2 \left[\pi_\infty\norm{A}_2^2+(1-\pi_\infty)\norm{A-BL}_2^2\right] 
\end{eqnarray}
Here Eqn.49 is obtained from the triangle inequality of spectral norm, Eqn.50 is obtained from the scalar multiplication of spectral norm, and Eqn.51 is obtained from the submultiplicative inequality of spectral norm. Moreover, $\pi_\infty = \left(\frac{p}{p+q}\right)$. Clearly, the necessary and sufficient condition for $\mathcal{T}_\infty$ to be a contraction is given by:
$\left[\left(\frac{p}{p+q}\right)\norm{A}_2^2+\left(\frac{q}{p+q}\right)\norm{A-BL}_2^2\right]<1$. By rearrangement, Eqn. \ref{vec_markov_iff_l2} of Theorem \ref{theorem16} is obtained. That concludes the proof. \hspace{14.5cm}$\blacksquare$

\subsection{Proof of Lemma \ref{lemma17}}
\label{proof17}
The stochastic dynamics based on the Markov switching random variable $\pmb \gamma_k$ is given by:
    \begin{eqnarray*}
        {\bf x}_{k+1} = {\pmb\Lambda}_k{\bf x}_k+(1-{\pmb\gamma}_k)BL{\bf e}_k+{\bf w}_k
    \end{eqnarray*}
    Where, ${\pmb\Lambda}_k = \left\{{\pmb\gamma}_kA+(1-{\pmb\gamma}_k)(A-BL)\right\}$ is the random state transition matrix. The conditional state covariance matrix given the sequence ${\pmb\gamma}_{0:k}$ is given by:
    \begin{eqnarray*}
        P_{k+1|{\pmb\gamma}_{0:k}}
        =  (1-{\pmb\gamma}_k)(A-BL)P_{k|{\pmb \gamma}_{0:k-1}}(A-BL)'\\
        + {\pmb\gamma}_k A P_{k|{\pmb \gamma}_{0:k-1}}A'
        + (1-{\pmb\gamma}_k)BL\mathbb{E}({\bf e}_k{\bf e}_k'|{\pmb\gamma}_{0:k-1})L'B'+W
    \end{eqnarray*}
    Since $m<n$, the matrix $BL$ is singular, the innovation contribution (differential entropy) of the quantization error vector to the state differential entropy is $-\infty$, resulting in the volume of the corresponding typical support set being zero. That concludes the proof. \hspace{12.2cm}$\blacksquare$

\subsection{Proof of Theorem \ref{theorem18}}
\label{proof18}
The proof is identical with the proof given in section \ref{proof9} for Theorem \ref{theorem9}. That concludes the proof. \hspace{2cm}$\blacksquare$

\end{document}